\documentclass[pdflatex,sn-mathphys-ay]{sn-jnl}

\usepackage[T1]{fontenc}
\usepackage[utf8]{inputenc}
\DeclareUnicodeCharacter{2122}{\texttrademark}

\usepackage{graphicx}
\usepackage{subcaption}
\usepackage{caption}
\usepackage{float}
\usepackage{placeins}
\usepackage{wrapfig}

\usepackage{amsmath}
\usepackage{amssymb}
\usepackage{tikz}
\usepackage{bm}
\usepackage{comment}

\usepackage{booktabs}
\usepackage{array}

\usepackage{ragged2e}
\usepackage{xcolor}
\usepackage{textcomp}
\usepackage{circledsteps}
\usepackage{pifont}
\usepackage[percent]{overpic}
\usepackage{adjustbox}

\usepackage[colorinlistoftodos,textwidth=25mm]{todonotes}

\providecommand*\CircledText[1]{}
\renewcommand*\CircledText[1]{%
  \tikz[baseline=(char.base)]{
    \node[
      shape=circle,
      draw,
      inner sep=0.4pt,
      line width=0.4pt,
      font=\scriptsize
    ] (char) {#1};
  }%
}

\newcommand{\ca}{\mathrm{Ca}}

\begin{document}

\title[Transparent 3D-Printed Flow Cells]{Transparent 3D-printed flow cells for porous media studies with proof-of-concept on two-phase flow-reversal asymmetry}

\author*[1]{\fnm{Yann} \sur{Dumay}}
\email{yann.dumay@ntnu.no}

\author[2]{\fnm{J. Kevin} \sur{Pierce}}

\author[1]{\fnm{Julie} \sur{Delhaie}}

\author*[1,3]{\fnm{Erika} \sur{Eiser}}

\affil*[1]{\orgdiv{Department of Physics, PoreLab}, \orgname{Norwegian University of Science and Technology (NTNU)}, \orgaddress{\city{Trondheim N-7491}, \country{Norway}}}

\affil[2]{\orgdiv{Departments of Geoscience and Physics, PoreLab, Njord Centre}, \orgname{University of Oslo}, \orgaddress{\city{Oslo N-0316}, \country{Norway}}}

\affil[3]{\orgdiv{Department of Physics}, \orgname{University of Cambridge}, \orgaddress{\city{Cambridge CB3 0HE}, \country{United Kingdom}}}
\email{erika.eiser@ntnu.no}
\abstract{
Porous media experiments commonly rely on geometry-controlled and optically transparent experimental flow cells.
Examples include studies of flow structure, solute mixing, multiphase flow, bacterial dynamics, and reactive transport.
Here, we provide a reproducible and open-source protocol to fabricate transparent monolithic porous flow cells by stereolithography 3D printing.
The protocol covers digital design, resin printing, internal flushing, and resin coating.
The flushing and coating steps are central to the workflow, enabling enclosed single-material cells to be cleared of uncured resin and made optically transparent for direct imaging.
The protocol produces optically transparent, millimetre-scale, and quasi-two-dimensional flow cells suitable for simultaneous pressure and image-based measurements.
As a proof of concept, we study flow-reversal asymmetry in steady oil--air two-phase flow in quasi-two-dimensional Hele-Shaw cells containing either triangular or cylindrical obstacles.
Directional differences are quantified from pressure measurements and from image-based active-width measurements.
The triangular geometry produces a clear directional response, whereas the cylindrical geometry remains symmetric within experimental uncertainty. 
These results demonstrate that our fabrication protocol can link digitally designed pore geometries to macroscopic pressure response and steady-state two-phase flow organisation.
By making the design files, scripts, and manufacturing workflow openly available, we aim to make 3D-printed flow cells readily adaptable to other experimental systems and research fields.
}


\maketitle

\section*{Article Highlights}
\begin{itemize}
    \item Resin coating smooths printed optical surfaces and improves transparency for direct flow imaging.
    \item Digital design enables flexible control of pore geometry for optical flow experiments.
    \item Asymmetric pores produce a measurable flow-reversal asymmetry in steady two-phase flow.
\end{itemize}

\section{Introduction}

Transport processes in porous media are central to numerous scientific problems with important applications, including carbon dioxide sequestration~\citep{Krevor2023,Heinemann2021}, solute mixing and chemical reactions~\citep{LeBorgneHeyman2026}, fuel-cell and battery technologies~\citep{ZamelLi2013,Mukherjee2011}, and microbial processes~\citep{JinSengupta2024,TeconOr2017}.
Experimental studies of transport processes often rely on simplified porous structures in which pore geometry, boundary conditions, and fluid forcing can be simultaneously controlled.
A major challenge is to fabricate such systems with well-defined geometrical, material, and optical properties, so that transport can be imaged quantitatively and related directly to the pore-scale geometry~\citep{Anbari2018}.

Two-phase flow provides a particularly challenging example because the presence of fluid--fluid interfaces introduces capillary forces, interface pinning and rearrangements, phase trapping, and changes in fluid connectivity within the pore space~\citep{JoekarNiasar2012,Feder2022}.
These pore-scale processes can strongly influence the macroscopic flow response, making it dependent not only on the imposed forcing and phase saturation but also on the spatial organisation and dynamics of the two fluid phases~\citep{Valavanides2018,SinhaEtAl2017}.
Consequently, even the relationship between macroscopic pressure gradient and flow rate can depart from the linear Darcy response and remains an active topic in steady-state two-phase flow~\citep{SinhaHansen2012,Hansen2026}.
Resolving these relationships experimentally requires model porous media with precisely defined pore geometry and sufficient optical access to quantify the evolving fluid configuration and relate it to the measured macroscopic flow response.

Different techniques have been developed to produce porous models with both geometric control and optical access, each with specific advantages and limitations. Soft lithography and etching techniques provide excellent optical transparency and geometric control on micro- and millimetre scales. These methods require specialised facilities, however, and are primarily suited to quasi-two-dimensional geometries~\citep{Gerami2019,qin2010soft,karadimitriou2012review,jahanbakhsh2020review}.

An alternative approach consists of placing granular materials between flat plates, forming a porous analogue of a Hele-Shaw cell~\citep{Lovoll2004,TallakstadPRE2009}. Such systems can span scales from millimetres to metres~\citep{TallakstadPRE2009}, but provide more limited control over the individual pore geometry.

Fully three-dimensional granular packings provide a more representative three-dimensional pore space on micrometre to millimetre scales. These can be imaged by combining refractive-index matching and confocal microscopy~\citep{Datta2014,DoNascimento2019} or planar laser-induced fluorescence~\citep{souzy2020,heyman2021,brodin2022}. In this case, geometric control is again limited, while the requirement for refractive-index matching strongly restricts the available solid--fluid combinations~\citep{Wright2017,souzy2020,heyman2021,brodin2022}.

X-ray and neutron-based tomographic imaging have also been used to study transport processes in opaque porous samples, including rock cores and bead packs
~\citep{raeini2014direct,silin2011microtomography,tekseth20244d}.
These methods can require specialised imaging facilities, long acquisition times, contrast agents, and advanced tomographic reconstruction, substantially increasing experimental complexity and
cost
~\citep{Wright2017,souzy2020,heyman2021,brodin2022}.

Taken together, existing methods to fabricate and image porous cells have successfully resolved diverse transport processes across a wide range of scales and applications. Nevertheless, established approaches generally require compromises between optical access, geometric control, dimensionality, and ease of implementation, motivating the development of complementary flow-cell fabrication methods.

Additive manufacturing offers an attractive route for porous model fabrication because it allows three-dimensional structures to be generated directly from digital models~\citep{PatsoukisDimou2022,Waheed2016,Au2016,Nielsen2020}.

Stereolithography has been explored for experimental porous-media models for more than a decade, including early devices designed for pore-scale visualisation of immiscible two-phase displacement~\citep{Crandall2008}.

In particular, stereolithography resin printing has recently been used to fabricate flow cells with prescribed obstacle shapes, porosities, pore sizes, and inlet/outlet geometries~\citep{Nielsen2020,Musgrove2022}, enabling studies of fluid kinematics and deformation~\citep{Pierce2026,Linga2026}, multiphase flow~\citep{VincentDospital2022, Khobaib2025,Maloy2021}, and reactive transport~\citep{Zhang2025}.
Stereolithography resin printing is a type of additive manufacturing whereby sequential layers of a three-dimensional object are formed by the selective curing of a liquid resin using ultraviolet light~\citep{Au2016}.
Stereolithography printing of porous models is challenging because newly printed parts contain uncured residual resin and possess inherent surface roughness that degrades optical transparency~\citep{Waheed2016}.

To fabricate optically transparent porous cells, residual resin must be removed from internal channels, and the printed optical surfaces must be processed to provide sufficient transparency.
Several post-processing approaches have been developed to address these challenges, including mechanical polishing and surface coating~\citep{Waheed2016,Kim2023}.
Another approach is to print an ``open'' porous structure to allow removal of residual resin, then to seal it between transparent glass or polymer plates with added liquid films to fill surface roughness~\citep{Kecili2020,Khobaib2025,Pierce2026,Zhang2025,VincentDospital2022,Linga2026}.
Such composite cells provide optical access, but they also introduce interfaces between the transparent plate and the printed structure, requiring bonding, alignment, or external compression, ultimately increasing the risks of leakage, deformation, and refractive-index discontinuities~\citep{Kecili2020}.
In addition, the use of multiple materials can introduce wettability contrasts that complicate the interpretation of multiphase flow studies~\citep{VahidDastjerdi2022}.
To avoid these complications, it is desirable to fabricate self-contained flow cells from a single material, i.e., using ``monolithic'' construction.

Here, we present a reproducible protocol for fabricating transparent monolithic porous flow cells by stereolithography 3D printing.
The protocol combines digital design, resin printing, internal flushing, and an external resin-coating step to produce enclosed single-material flow cells suitable for direct optical imaging.
These digitally controlled porous geometries provide advantages when compared to bead-pack-based flow cells, for which the pore-scale configuration is difficult to reproduce and control~\citep{Maier2008}.

By making the design files, scripts, and fabrication workflow openly available, we provide a practical route for researchers to adapt 3D-printed flow cells to different experimental applications using consumer-grade stereolithography printers and commonly available equipment~\citep{Dumay2026Zenodo}.

As a proof of concept, we investigate steady-state oil--air two-phase flow in quasi-two-dimensional Hele-Shaw cells.
Steady-state two-phase flow provides a demanding test case because the macroscopic pressure response depends not only on the imposed flow rate and phase saturation, but also on the pore-scale organisation and rearrangement of fluid--fluid interfaces and active flow pathways~\citep{deGennes2004,TallakstadPRL2009,SinhaEtAl2017,Valavanides2018,Feder2022}.
We study flow-reversal asymmetry in cells containing triangular or cylindrical obstacles by comparing steady-state two-phase flow in opposite directions at the same imposed total flow rate.
We test whether reversing the flow direction alters either the pressure drop across the cell or the spatial organisation of the air phase, quantities that are known to depend on fluid configuration and interface dynamics in steady-state two-phase flow~\citep{TallakstadPRL2009,SinhaEtAl2017}.
We find enhanced flow-reversal asymmetry in porous media composed of triangular obstacles compared with cylindrical ones, demonstrating a geometry-dependent flow-reversal response in steady-state two-phase flow and illustrating one of the research possibilities enabled by the fabrication protocol.

\section{Protocol for flow-cell fabrication}
\label{sec:protocol}
The protocol described here provides a general framework for fabricating transparent monolithic porous flow cells by stereolithography 3D printing. Because the internal geometry is defined digitally, the approach can be adapted to a wide range of porous structures by varying parameters such as obstacle shape, pore size, porosity, spatial arrangement, and inlet/outlet configuration.

The manufacturing workflow consists of three main steps: digital design and 3D printing of the monolithic cell, flushing of the internal channels to remove residual uncured resin, and external resin coating to improve optical transparency.

A complete list of the materials and equipment required to reproduce the fabrication protocol, including manufacturers and product references, is provided in the accompanying Zenodo repository~\citep{Dumay2026Zenodo}.

\subsection{Geometry and 3D printing of the flow cells}

The flow-cell geometries are generated in FreeCAD~0.20.2 using the Jupyter notebook \texttt{FreeCAD\_obstacles\_generator.ipynb}~\citep{Dumay2026Zenodo}. The notebook creates the porous structure directly within the flow-cell design from user-defined geometrical parameters, including chamber dimensions, pillar shape, pillar size, target porosity, placement type, and orientation. Because these parameters can be modified independently, the same workflow can be adapted to a wide range of porous structures and experimental configurations.

For the proof-of-concept experiments presented here, we fabricate Hele-Shaw-type quasi-two-dimensional porous cells with a fixed gap height of $h=1\,\mathrm{mm}$, following the common use of thin-gap model porous media for optical studies of two-phase flow~\citep{TallakstadPRL2009}. Each chamber has dimensions of $30 \times 85 \times 1\,\mathrm{mm}^3$ and contains one inlet and one outlet.

A buffer region is included between each inlet or outlet port and the patterned porous region. These buffers allow the two phases to redistribute across the cell width before entering the porous structure and reduce the direct influence of the inlet and outlet geometry on the observed flow patterns.

For the porous geometries used in the proof of concept, the notebook is used to generate regular arrays of triangular and cylindrical pillars with a designed porosity of $\phi=0.85$. In the following, we refer to the two types of cells as T85 and C85, respectively. A blank reference cell, BL, containing only a small number of spacers is also fabricated, with $\phi\approx1$. 

The same notebook can additionally generate randomised porous structures by varying pillar positions and/or in-plane orientations, making the scripted design procedure directly reproducible and suitable for systematic variation of geometrical parameters.

\begin{figure}[H]
\centering

\begin{subfigure}[t]{0.495\textwidth}
    \centering
    \includegraphics[
        width=\textwidth,
        keepaspectratio
    ]{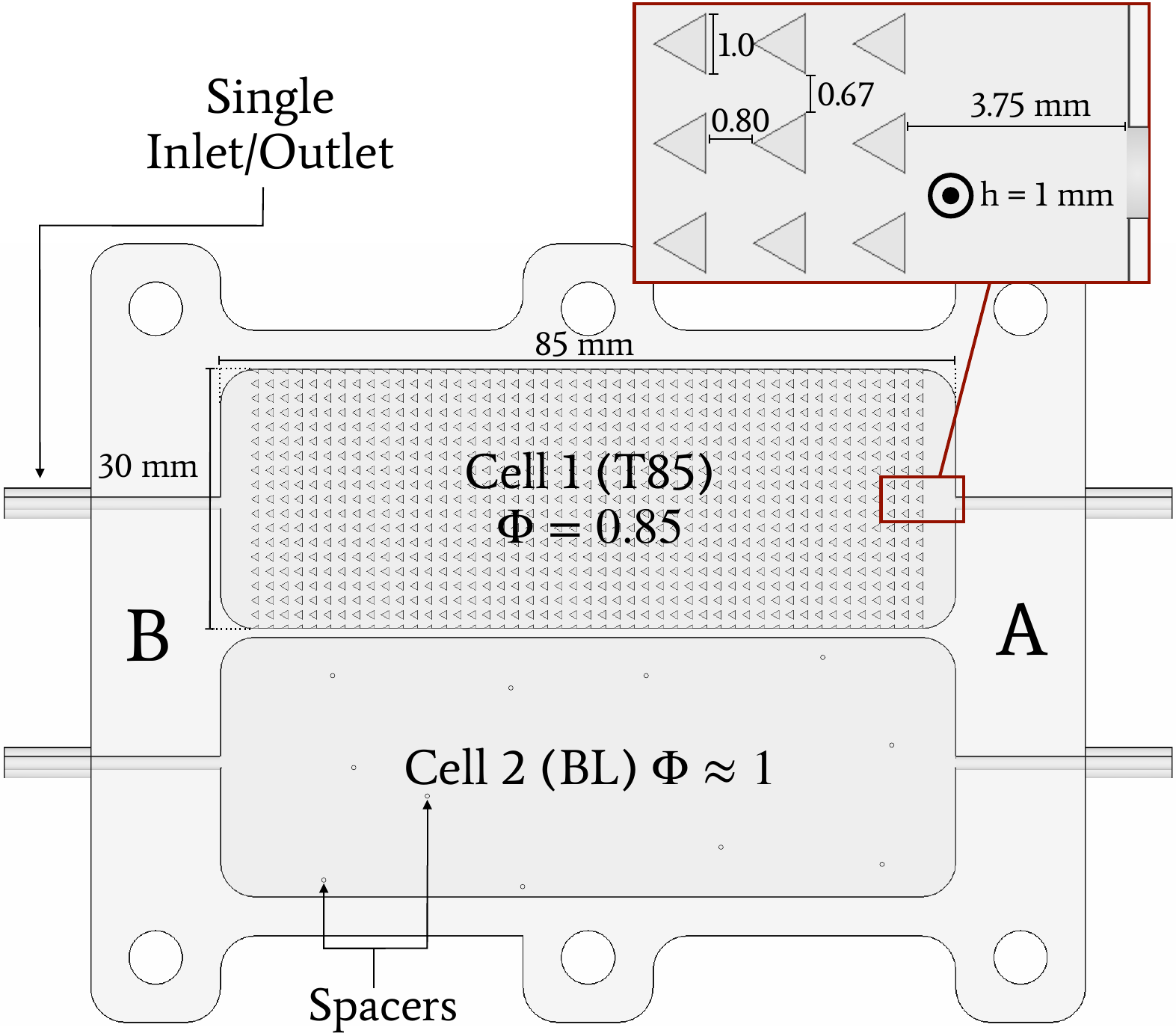}
    \captionsetup{justification=centering}
    \caption{}
    \label{fig:sample_struct_a}
\end{subfigure}
\hfill
\begin{subfigure}[t]{0.495\textwidth}
    \centering
    \includegraphics[
        width=\textwidth,
        keepaspectratio
    ]{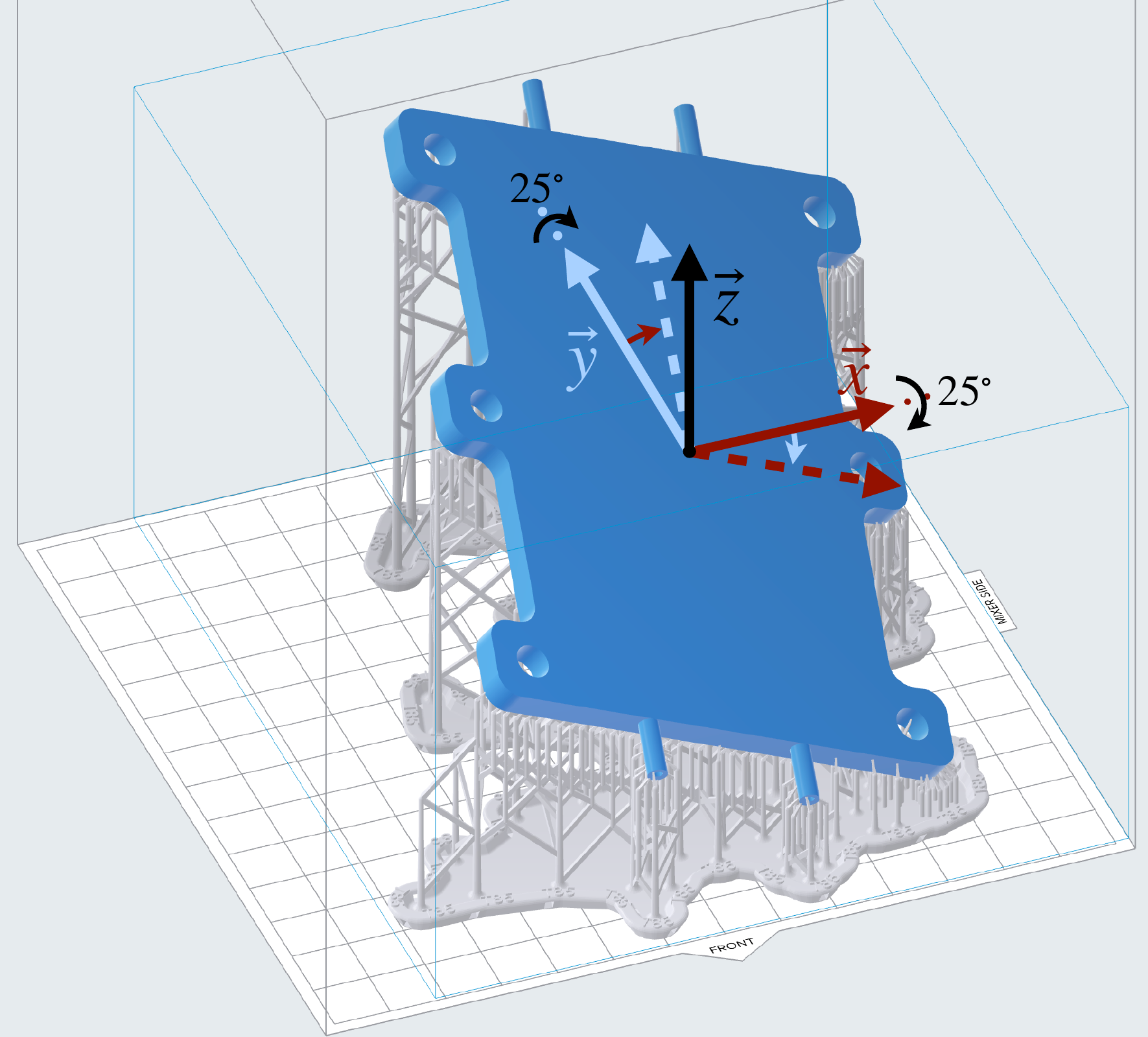}
    \captionsetup{justification=centering}
    \caption{}
    \label{fig:sample_struct_b}
\end{subfigure}

\caption{\justifying
Design and printing setup of the 3D-printed flow cells.
(a) Schematic of the 3D-printed flow-cell geometry with the triangular
porous matrix T85 and the BL reference cell described in
Sec.~\ref{sec:proof}.
(b) Printing orientation used for fabrication, illustrating the tilted
placement of the cell and the support structures generated during
printing.
}
\label{fig:sample_struct}
\end{figure}

Representative flow-cell designs and the printing orientation used for fabrication are shown in Fig.~\ref{fig:sample_struct}.

Flow cells are fabricated using Clear Resin V4.1 (Formlabs, USA) on a
Form~3B+ stereolithography printer (Formlabs, USA), with a nominal XY resolution of $25\,\mu\mathrm{m}$. A layer thickness of $25\,\mu\mathrm{m}$ is used, corresponding to the selected Z resolution.
The resulting flow-cell geometry is exported from FreeCAD as an STL file and imported into PreForm~3.46.0.497, the slicing software provided by Formlabs.

The parts are oriented in PreForm with tilt angles of $X=25^\circ$ and $Y=25^\circ$ (Fig.~\ref{fig:sample_struct}(b)).
This orientation facilitates drainage of uncured resin from the enclosed cell through the inlet and outlet ports during and after printing. External support structures are generated using a full raft, with a touchpoint density of $0.85$ and a touchpoint size of $0.20\,\mathrm{mm}$.
No internal supports are added within the flow channels or porous region.
After printing, the external supports are removed manually, taking care not to damage the inlet or outlet ports.

The same Clear Resin V4.1 is used for the post-printing coating step described below.
Although the present protocol is demonstrated using this specific printer and resin, the workflow is not inherently restricted to this combination.
Other stereolithography printers and transparent resins could in principle be used, although printing accuracy, surface quality, optical transparency, and material properties will depend on the selected printing system and resin.

The corresponding STL, STEP, and FreeCAD design files are available on Zenodo~\citep{Dumay2026Zenodo}.

\FloatBarrier
\subsection{Flushing of residual resin}

After printing, the external surfaces of the cell are washed by immersing the printed part in an isopropanol bath (Sigma-Aldrich, product no. 1.09634.6025) and sonicating it for $10\,\mathrm{min}$.

\begin{figure}[H]
    \centering
    \includegraphics[
        width=0.95\textwidth,
        keepaspectratio
    ]{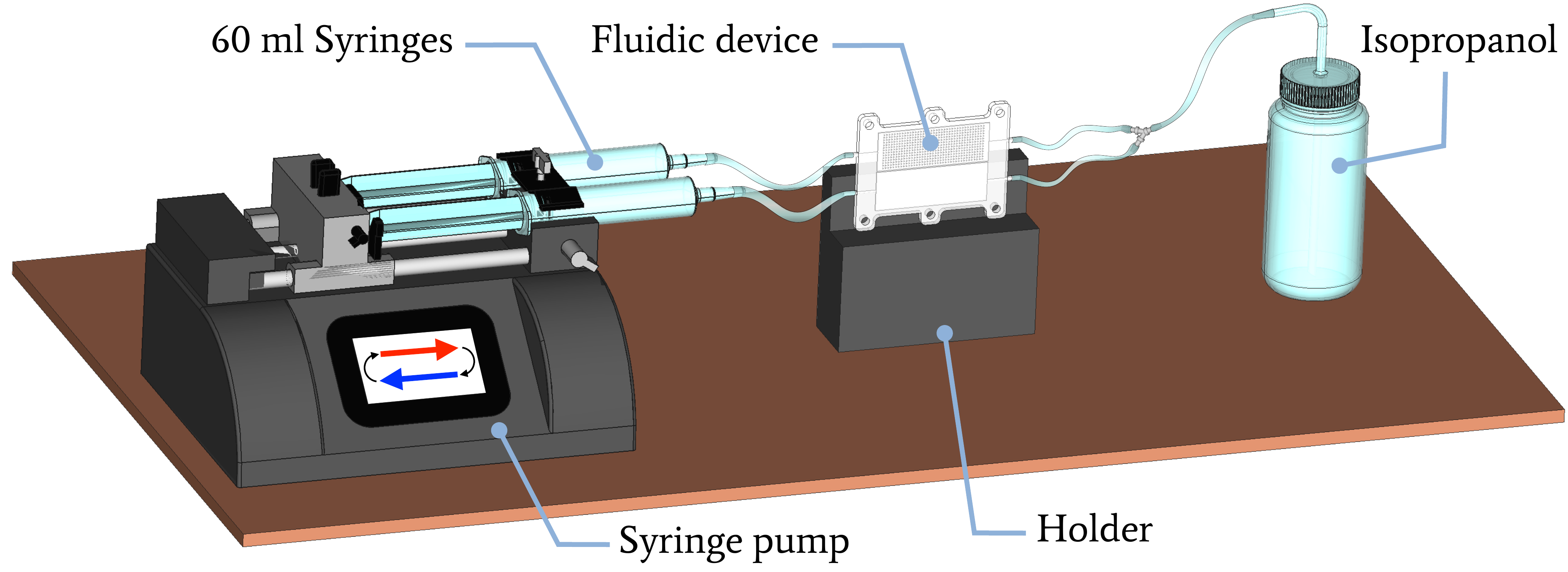}

    \caption{\justifying
Illustration of flushing setup used to remove residual uncured resin from the 3D-printed flow cell.
The device is connected between a programmable syringe pump and an isopropanol reservoir, enabling repeated infuse/withdraw cycles through the internal channels.
A dedicated holder keeps the fluidic device in a near-vertical position, facilitating drainage and removal of trapped resin.
}

    \label{fig:flush}
\end{figure}

This first washing step does not remove uncured residual resin inside the cell.
Therefore, the inlet and outlet channels, as well as the internal chamber, need to be flushed with additional isopropanol immediately after the first washing step, using the setup shown in Fig.~\ref{fig:flush}.
For this step, one end of the fluidic device is connected to a programmable syringe pump using flexible Tygon tubing, while the other end is connected to a squeeze bottle filled with isopropanol. The cap of the squeeze bottle is drilled with a $3\,\mathrm{mm}$ vent hole to prevent negative pressure from building up inside the bottle during withdrawal. The tubing used for flushing is Tygon E-3603 tubing, with an inner diameter of $3.18\,\mathrm{mm}$ and an outer diameter of $6.36\,\mathrm{mm}$, product code ACF00007-S, supplied by Saint-Gobain (France).

The syringe pump is then used to run repeated infuse/withdraw cycles, which force isopropanol back and forth through the full internal volume of the device. 
The flushing is first performed at a low flow rate ($300\,\mathrm{mL\,h^{-1}}$) to avoid excessive pressure build-up inside the freshly printed cell. The flow rate is then increased to $1200\,\mathrm{mL\,h^{-1}}$ to accelerate the process.

Halfway through the flushing procedure, the device is rotated by $180^\circ$, so that the side initially facing upward is placed downward, and vice versa.
This inversion helps to mobilise resin trapped by gravity or held in local pockets within the porous structure.
The duration and flow rate used during flushing depend on the geometry of the device, including the cell length, width, thickness, pore volume, pore connectivity, porosity, and number and position of inlets and outlets.
More open structures can therefore be cleared faster, whereas denser or poorly connected porous matrices may require longer flushing times or additional vents.
For the T85 and C85 cells with a chamber size of $30 \times 85 \times 1\,\mathrm{mm}^3$ and a single inlet and outlet, the residual resin is typically fully removed after about $10\,\mathrm{min}$ of flushing.

It is important to perform the flushing step in a darkened environment or under reduced ambient light to limit premature curing of resin dissolved in the isopropanol. 
If the dissolved resin cures during flushing, solid clusters can form inside the tubing or within the device, which may clog the pores or alter the geometry of the matrix.

After flushing, the cell is dried by flowing air continuously through the internal channels for approximately $15\,\mathrm{min}$ to remove residual isopropanol. 
For more intricate porous structures, lower-porosity designs, or cells with smaller pore sizes, this drying step can be extended to several hours to ensure complete solvent removal from the internal volume.
The device is then inspected visually to check for blocked pores, trapped resin, cracks, leaks, or damage on the inlet and outlet ports. 
At this stage, the printed part is already transparent enough to verify whether the internal structure has been successfully cleared, but the optical quality is still limited by surface roughness of the printed surfaces.

\subsection{Surface coating for optical transparency}

\begin{figure}[H]
\centering
\includegraphics[
width=0.95\textwidth,
keepaspectratio
]{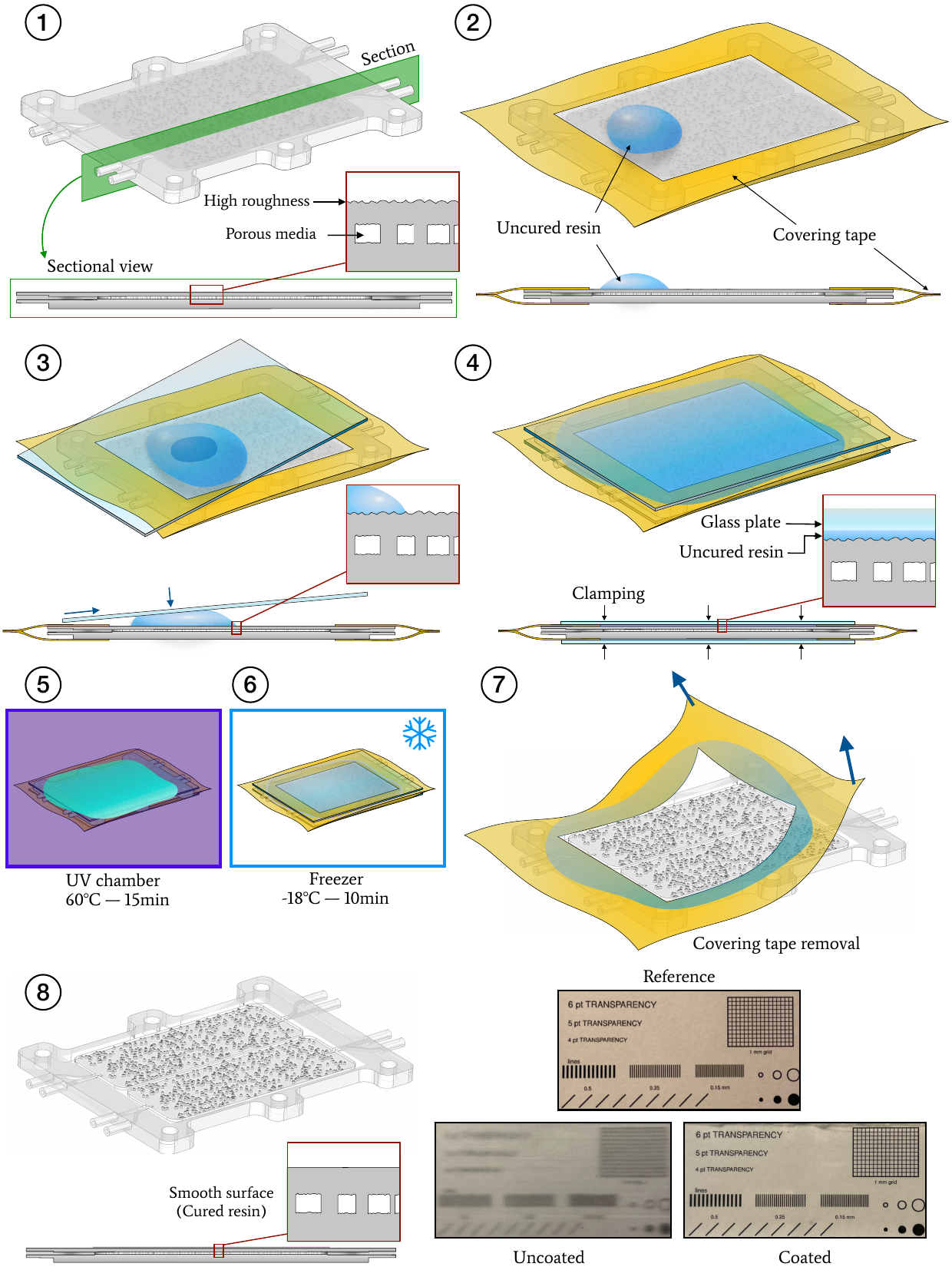}
\caption{\justifying
Resin-coating procedure used to improve the optical transparency of the 3D-printed flow cells.
\protect\CircledText{1} After printing and flushing, the surface remains rough and optically scattering.
\protect\CircledText{2} The imaging region is delimited with tape, and uncured Clear Resin V4.1 is deposited on the surface.
\protect\CircledText{3} A glass plate spreads the resin over the imaging window.
\protect\CircledText{4} The assembly is clamped to keep the interface flat.
\protect\CircledText{5} The coating is cured at $60^\circ\mathrm{C}$ for $15\,\mathrm{min}$.
\protect\CircledText{6} The cell is cooled at approximately $-18^\circ\mathrm{C}$ for $10$--$15\,\mathrm{min}$.
\protect\CircledText{7} The tape and glass plate are removed.
\protect\CircledText{8} The cured coating forms a smooth resin layer that improves transparency while preserving a monolithic, single-material device. 
The optical comparison shows the improvement relative to the uncoated surfaces.
}
\label{fig:Coating}
\end{figure}

To improve optical transparency, we apply a thin resin coating to the external optical surfaces of the printed cell, as illustrated in Fig.~\ref{fig:Coating}.
The coating is made from the same Clear Resin V4.1 as the printed cell.
Using the same resin avoids introducing an additional refractive-index mismatch at the optical surfaces.
The coating primarily acts by filling small printing-induced surface irregularities, thereby producing smoother optical surfaces and enabling clearer imaging of the two-phase flow.

The coating procedure is as follows:
After the flushing and drying steps, the top and bottom imaging windows of the printed cell are delimited using a covering tape (3M™ Multi-Use Duct Tape 2900, 3M, USA; Fig.~\ref{fig:Coating}\,\CircledText{2}).
Before depositing uncured Clear Resin V4.1 on the cell windows, we transfer the resin into an opaque $60\,\mathrm{mL}$ test tube to limit exposure to ambient light, thus reducing the risk of premature curing during handling.
The resin is then degassed in a desiccator for $60\,\mathrm{min}$ to remove entrained air and reduce bubble formation during coating.
A small volume of the degassed resin is deposited slowly onto the cell, preferentially from one side of the imaging window rather than directly at the centre (Fig.~\ref{fig:Coating}\,\CircledText{2}).

A clean borosilicate glass plate, $3\,\mathrm{mm}$ thick, is then brought into contact with the resin at a small angle and slowly lowered onto the surface (Fig.~\ref{fig:Coating}\,\CircledText{3} and~\CircledText{4}). This deposition step is the most critical part of the coating procedure, because air bubbles trapped between the resin layer and the glass plate can degrade the optical quality of the final device.
Using a relatively thick glass plate makes it easier to handle and reduces the risk of breakage during removal after curing.

After the first side has been coated, the cell is flipped and the same procedure is repeated on the opposite side.
After both glass plates are in place, plastic clamps are used to hold the glass plates against the cell and maintain a flat interface during curing.
Although the standard curing procedure for Clear Resin involves preheating the printed part in the UV chamber, this step is avoided here to prevent the uncured coating resin from flowing out of the imaging window before curing.
Instead, the UV chamber is first preheated to $60^\circ\mathrm{C}$ before the cell with the clamped glass plates is added (Fig.~\ref{fig:Coating}\,\CircledText{5}).
The curing step is performed at $60^\circ\mathrm{C}$ for $15\,\mathrm{min}$.
When clamps overlap the region of interest, they are repositioned halfway through curing to allow the resin underneath them to cure properly.

After curing, the clamps are removed and the coated cell is placed in a freezer at approximately $-18^\circ\mathrm{C}$ for $10$--$15\,\mathrm{min}$ (Fig.~\ref{fig:Coating}\,\CircledText{6}).
This cooling step facilitates removal of the glass plates by exploiting the different thermal contraction of the glass and the cured resin.
In some cases, this contraction is sufficient for the glass plate to detach spontaneously from the resin surface.
During curing and cooling, the glass plates also help keep the entire device flat, limiting bending or warping caused by thermal stresses in the device.
Finally, the covering tape and glass plates are removed carefully (Fig.~\ref{fig:Coating}\,\CircledText{7}), leaving a smooth cured resin layer on both optical surfaces of the printed cell (Fig.~\ref{fig:Coating}\,\CircledText{8}).

\section{Proof of Concept: Flow-Reversal Asymmetry in Two-Phase Flow} \label{sec:proof}

To demonstrate the functionality of our cell-fabrication protocol, we investigate steady two-phase flow using oil and air in both flow directions, AB ($A\rightarrow B$) and BA ($B\rightarrow A$).

For Newtonian single-phase flow in the creeping-flow regime, the macroscopic pressure--flow relation is reciprocal under flow reversal: reversing the pressure gradient reverses the flow while leaving the hydraulic resistance unchanged~\citep{Bear1972}. Direction-dependent resistance has nevertheless been engineered at low Reynolds numbers using non-Newtonian fluids; notably, periodic arrays of triangular obstacles have been used to rectify viscoelastic polymer flow through an anisotropic porous medium~\citep{Kawale2019}. In steady two-phase flow, fluid--fluid interfaces introduce additional capillary thresholds, trapping, phase rearrangements, and history dependence~\citep[e.g.,][]{JoekarNiasar2013,Schluter2016}.

Previous studies have investigated time-reversal invariance in stationary two-phase flow through Onsager reciprocal relations and temporal correlations between the fluctuations of the two-phase fluxes~\citep{Winkler2020,Sales2022}.
Here, we consider a distinct macroscopic question: whether an established steady two-phase flow exhibits (i) the same pressure response and (ii) the same spatial phase organisation when the direction of the imposed mean flow is reversed while its magnitude and phase-flow ratio are kept unchanged. We refer to this property as flow-reversal symmetry.

To evaluate this question, we establish steady oil--air flows in the fabricated porous cells and monitor both the pressure drop across the cell and the spatial distribution of the fluid phases obtained from video analysis. Using these complementary measurements, we quantify the degree of flow-reversal asymmetry in T85 and C85, using BL (blank) as a reference cell.
The T85 and BL geometries are illustrated in Fig.~\ref{fig:sample_struct}(a).

\subsection{Experimental setup}

\begin{figure}[htbp]
    \centering
    \includegraphics[width=0.75\textwidth]
    {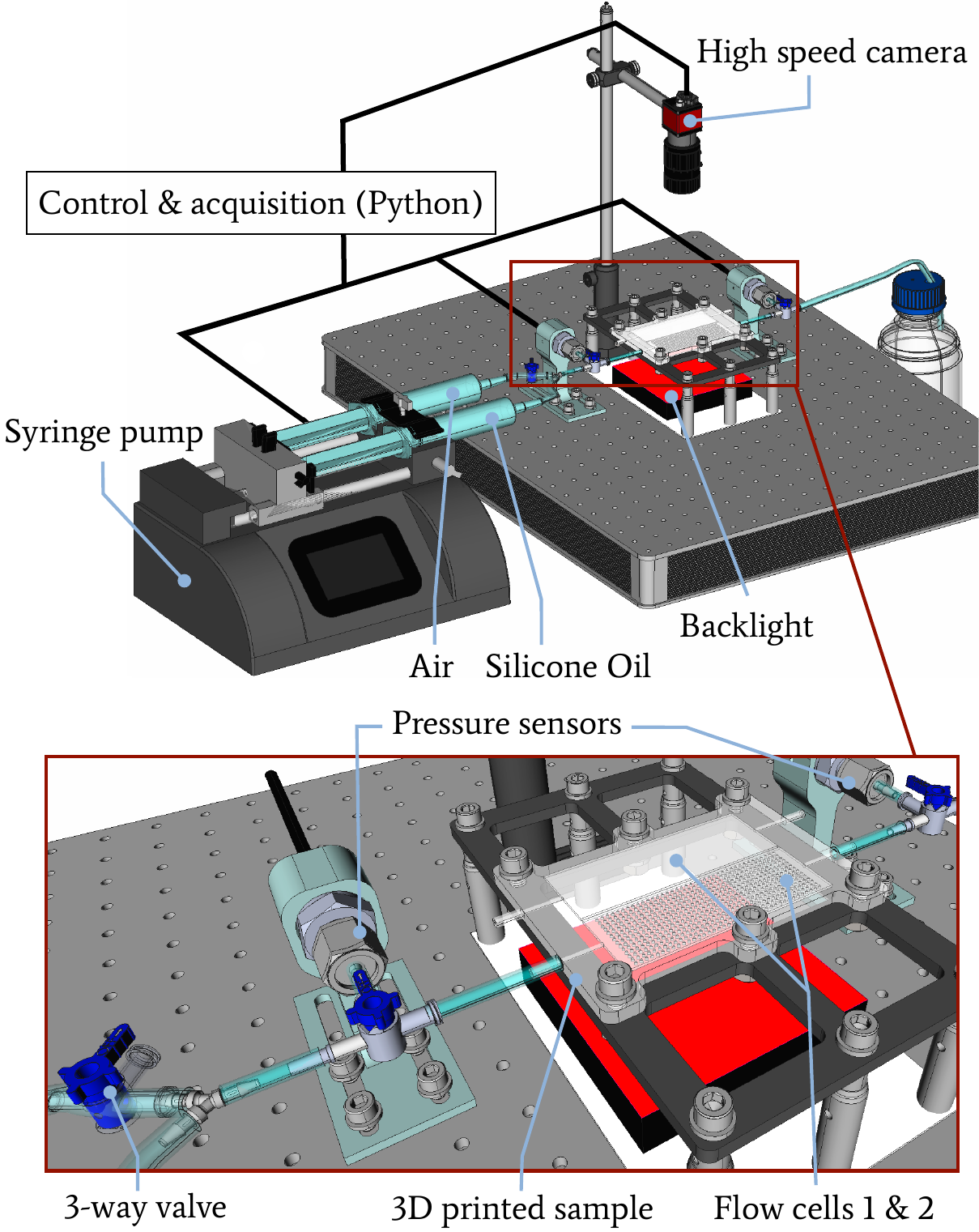}
    \caption{Experimental setup for pressure measurements and optical imaging during two-phase flow. Air and silicone oil merge upstream of the flow cell. Sensors near the inlet and outlet measure the pressure drop, while a monochrome camera records the phase distribution under diffuse red back-illumination.}
    \label{fig:setup}
\end{figure}

Air and silicone oil with a dynamic viscosity of $100\,\mathrm{mPa\,s}$ (Sigma-Aldrich, USA; product no.~378364) are co-injected using a PHD Ultra syringe pump (Harvard Apparatus, USA). The two streams merge into a single channel upstream of the flow cell (Fig.~\ref{fig:setup}). 

The pressure drop across the cell is obtained from the difference between two EZO-PRS pressure sensors (Atlas Scientific, USA) positioned near the inlet and outlet. Fluid-phase distributions are recorded using a monochrome camera (GS3-U3-23S6M-C, FLIR Integrated Imaging Solutions, Canada) equipped with a $50\,\mathrm{mm}$ lens (\#59-873, Edmund Optics, USA). The cell is placed between the camera and a diffuse red backlight with an illuminated area of $10 \times 10\,\mathrm{cm}^2$ (Nanoleaf Blocks, Nanoleaf, Canada). Red illumination was used to minimise resin autofluorescence, which is
stronger under shorter-wavelength excitation in Formlabs Clear Resin~\citep{Christopher2024}. All experiments are conducted in a darkened enclosure to minimise ambient-light fluctuations.

Custom 3D-printed cell holders, pressure-sensor holders, and sensor-to-barb adapters support and connect the apparatus. Their design files are available in the accompanying repository~\citep{Dumay2026Zenodo}.

\subsection{Experimental protocol and data acquisition}

Each cell is tested in both flow directions, AB and BA, at total imposed flow rates ranging from $40$ to $1000\,\mathrm{mL\,h^{-1}}$. The flow is divided equally between oil and air, $Q_o = Q_g = Q/2$.
Before each experiment, the cell is fully saturated with oil by closing the valve downstream of the air syringe and injecting oil using the syringe-pump controls. Subsequent pump operation, pressure recording, and camera triggering are automated using the Python notebook \texttt{Acquisition\_Pump\_Camera\_Prs-Sensors.ipynb}~\citep{Dumay2026Zenodo}.

At the start of each run, the prescribed flow rate is imposed and pressure recording begins at $1\,\mathrm{Hz}$. Both phases are co-injected and the system is allowed to evolve for
$150\,\mathrm{s}$ before image acquisition begins.

We identify steady state as a regime in which the pressure drop and air saturation fluctuate around approximately constant mean values without a persistent drift. Preliminary long-duration measurements were used to assess the approach to this regime and showed that the largest initial variations in the pressure response occurred during the first part of the experiment. Based on these measurements, a fixed waiting time of $150\,\mathrm{s}$ is used for all investigated flow rates and geometries.

Each run lasts $300\,\mathrm{s}$ in total. Pressure is recorded throughout the experiment, with the first $150\,\mathrm{s}$ used as a waiting period and images acquired during the final $150\,\mathrm{s}$. Frame rates of $10\,\mathrm{fps}$ and $20\,\mathrm{fps}$ are used below $300\,\mathrm{mL\,h^{-1}}$ and at or above $300\,\mathrm{mL\,h^{-1}}$, respectively. The mean steady-state pressure drop $\Delta P$ is calculated by averaging the pressure signal over the final $150\,\mathrm{s}$ of each run.
For each cell geometry, flow rate, and flow direction, the experiment is repeated four times.
The same protocol is used in both directions to compare the pressure response and phase distribution under flow reversal.

\subsection{Flow-regime characterisation}

The Reynolds and capillary numbers for phase $\alpha$, where $\alpha=o$ denotes oil and $\alpha=g$ denotes gas, are defined as
\begin{equation}
\mathrm{Re}_{\alpha}
= \frac{\rho_{\alpha} u_{\alpha} d}{\mu_{\alpha}},
\qquad
\mathrm{Ca}_{\alpha}
= \frac{\mu_{\alpha} u_{\alpha}}{\sigma}.
\end{equation}
Here, $\rho_{\alpha}$ and $\mu_{\alpha}$ are the mass densities and dynamic viscosities, $\sigma$ is the oil--air interfacial tension, and $d$ is a characteristic pore size.
The mean phase velocity is estimated as
\begin{equation}
u_{\alpha} = \frac{Q_{\alpha}}{\phi A S_{\alpha}},
\end{equation}
where $\phi$ is the porosity, $A$ is the cell cross-sectional area normal to flow, and $S_{\alpha}$ is the phase saturation.

For a porous region of area $A_p$ containing $N$ obstacles with equivalent-area radius $r_{\mathrm{obs}}$, the characteristic pore size is estimated as
\begin{equation}
d = \sqrt{\frac{A_p}{\pi N}} - r_{\mathrm{obs}}.
\end{equation}
Here, $r_{\mathrm{obs}}$ is defined as the radius of a circle having the same area as the obstacle. This length scale therefore accounts for both obstacle size and number density.

\begin{figure}[htbp]
    \centering
    \includegraphics[width=0.49\textwidth]
    {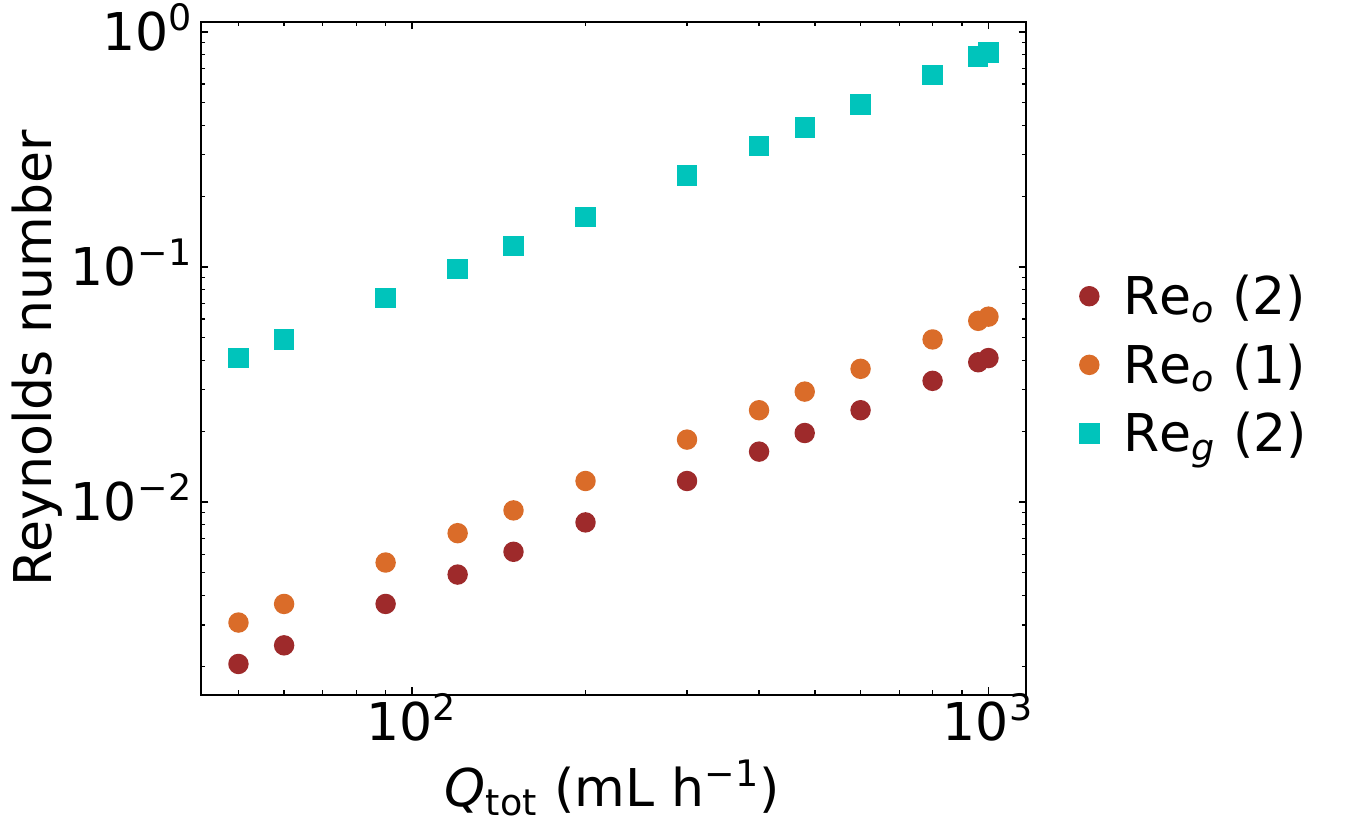}
    \hfill
    \includegraphics[width=0.49\textwidth]
    {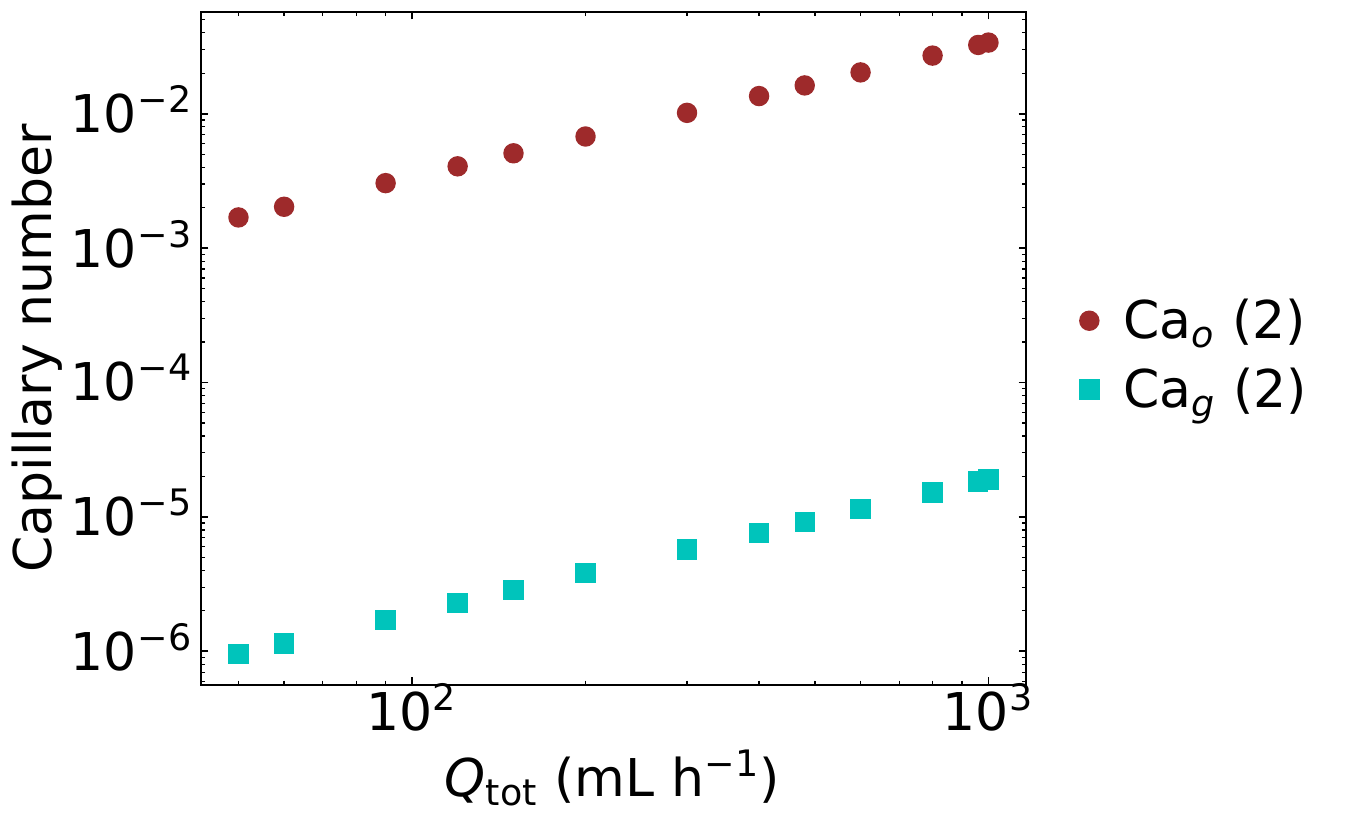}
    \caption{Estimated Reynolds numbers (left) and capillary numbers
    (right) as functions of total volumetric flow rate. Subscripts
    $o$ and $g$ denote oil and gas; labels $(1)$ and $(2)$ indicate
    single-phase oil flow and two-phase oil--air flow, respectively.
    Calculations are provided in \texttt{Plot\_Re\_Ca.ipynb}
    in the accompanying repository~\citep{Dumay2026Zenodo}.}
    \label{fig:ReCa}
\end{figure}

Fig.~\ref{fig:ReCa} summarises the estimated Reynolds and capillary numbers over the investigated flow-rate range.
The Reynolds and capillary numbers were calculated using \texttt{Plot\_Re\_Ca.ipynb}~\citep{Dumay2026Zenodo}.
The oil Reynolds number remains well below unity, while the gas Reynolds number remains of order unity or below, indicating weak inertial effects over the investigated range.

The wetting properties of the oil on the 3D-printed solid were characterised with a drop-shape analyser, using $0.5\,\mu\mathrm{L}$ sessile drops.
Observed contact angles decrease towards zero as a power-law in time, consistent with the silicone oil being fully wetting on the 3D-print resin.
Because oil is the wetting phase, the oil-based capillary number is used to summarise the balance between viscous and capillary effects.
The range of flow rates studied here gives $10^{-3}\lesssim \ca_o \lesssim 1.5 \times 10^{-2}$.

\FloatBarrier
\subsection{Air phase segmentation}
\label{Image_analysis}
\FloatBarrier

\begin{figure}
\centering

\begin{minipage}[t]{0.04\textwidth}
    \vspace{0pt}
    \textbf{(a)}
\end{minipage}%
\begin{minipage}[t]{0.82\textwidth}
    \vspace{0pt}
    \centering
    \includegraphics[
        width=\textwidth,
        keepaspectratio
    ]{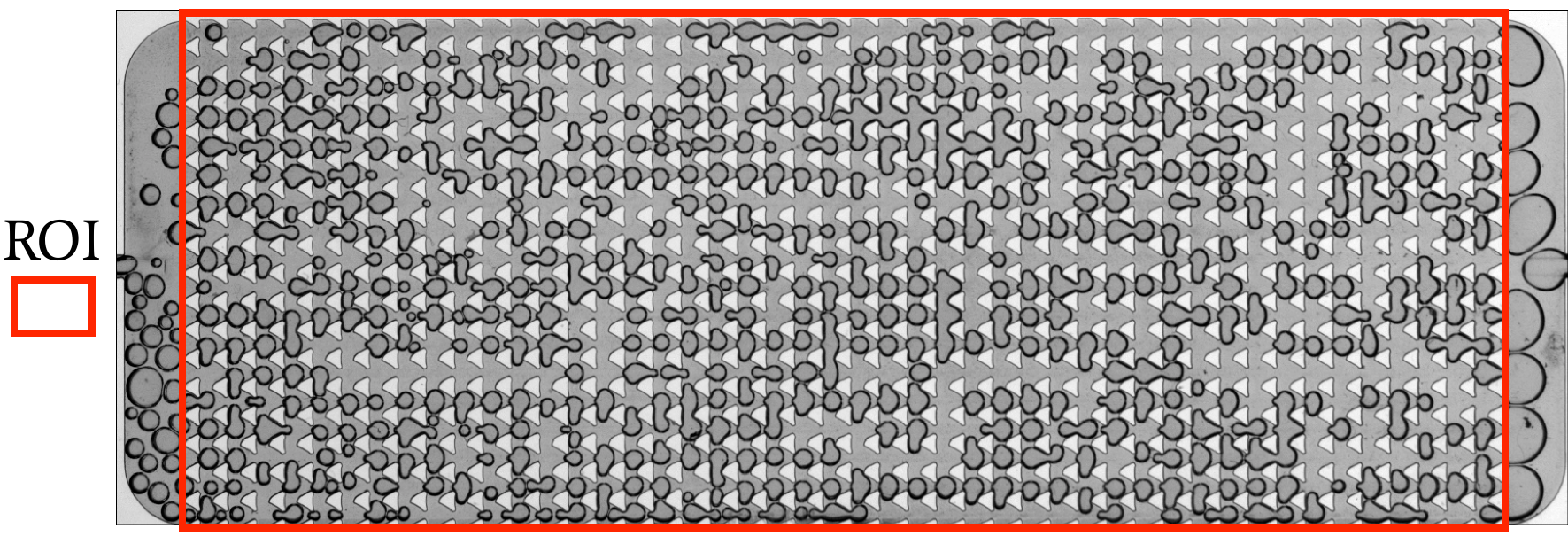}
\end{minipage}

\begin{minipage}[t]{0.04\textwidth}
    \vspace{0pt}
    \textbf{(b)}
\end{minipage}%
\begin{minipage}[t]{0.82\textwidth}
    \vspace{0pt}
    \centering
    \includegraphics[
        width=\textwidth,
        keepaspectratio
    ]{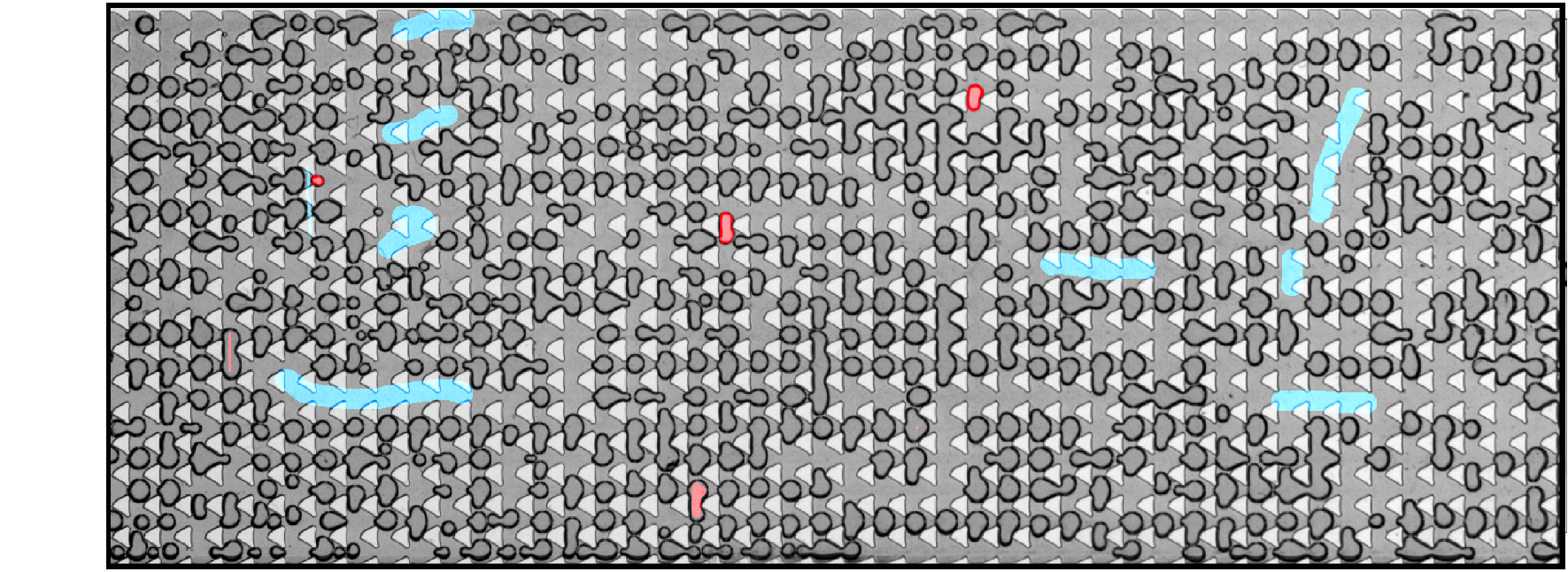}
\end{minipage}

\begin{minipage}[t]{0.04\textwidth}
    \vspace{0pt}
    \textbf{(c)}
\end{minipage}%
\begin{minipage}[t]{0.82\textwidth}
    \vspace{0pt}
    \centering
    \includegraphics[
        width=\textwidth,
        keepaspectratio
    ]{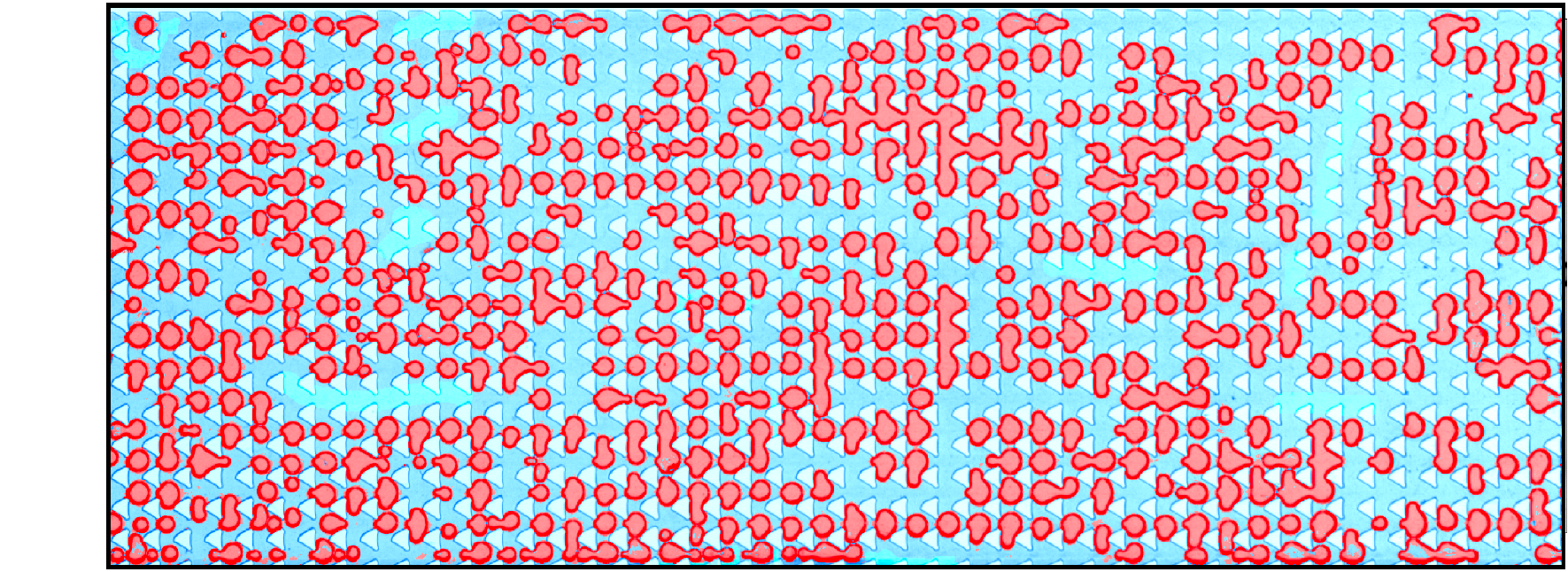}
\end{minipage}

\caption{\justifying
Image-processing workflow used to train the LabKit classifier.
(a) Representative raw frame extracted from the image stack, with the red rectangle indicating the region of interest retained for the analysis.
(b) Cropped frame opened in LabKit, where representative regions are manually labelled as foreground (red) and background (blue) using the drawing tool.
(c) Segmentation preview after classifier training, showing the air phase (red) separated from the oil and solid matrix (blue).
Once the classifier result is satisfactory, the trained classifier is saved and applied to the corresponding image stack using the Python batch-processing script.
}
\label{fig:LabKit}
\end{figure}

\begin{figure}
\centering

\begin{minipage}[t]{0.04\textwidth}
    \vspace{0pt}
    \textbf{(a)}
\end{minipage}%
\begin{minipage}[t]{0.85\textwidth}
    \vspace{0pt}
    \centering
    \includegraphics[
        width=\textwidth,
        keepaspectratio
    ]{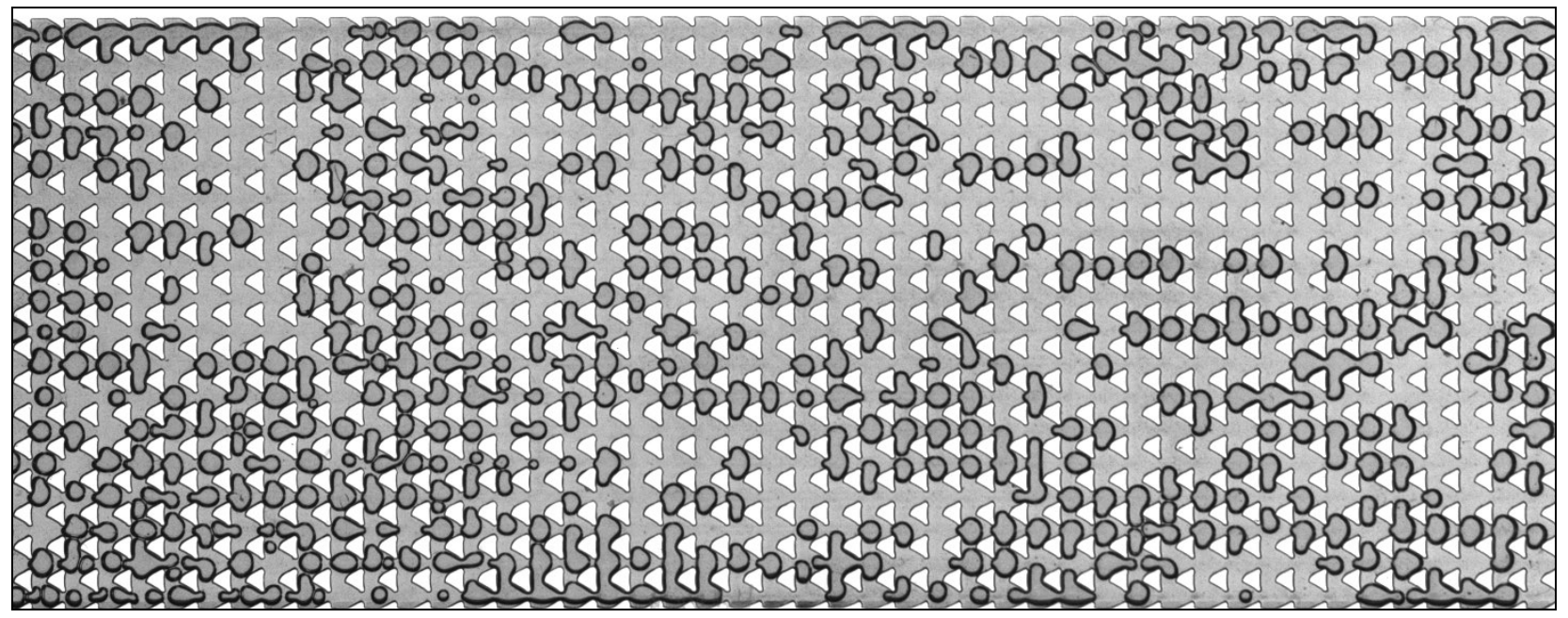}
\end{minipage}

\begin{minipage}[t]{0.04\textwidth}
    \vspace{0pt}
    \textbf{(b)}
\end{minipage}%
\begin{minipage}[t]{0.85\textwidth}
    \vspace{0pt}
    \centering
    \includegraphics[
        width=\textwidth,
        keepaspectratio
    ]{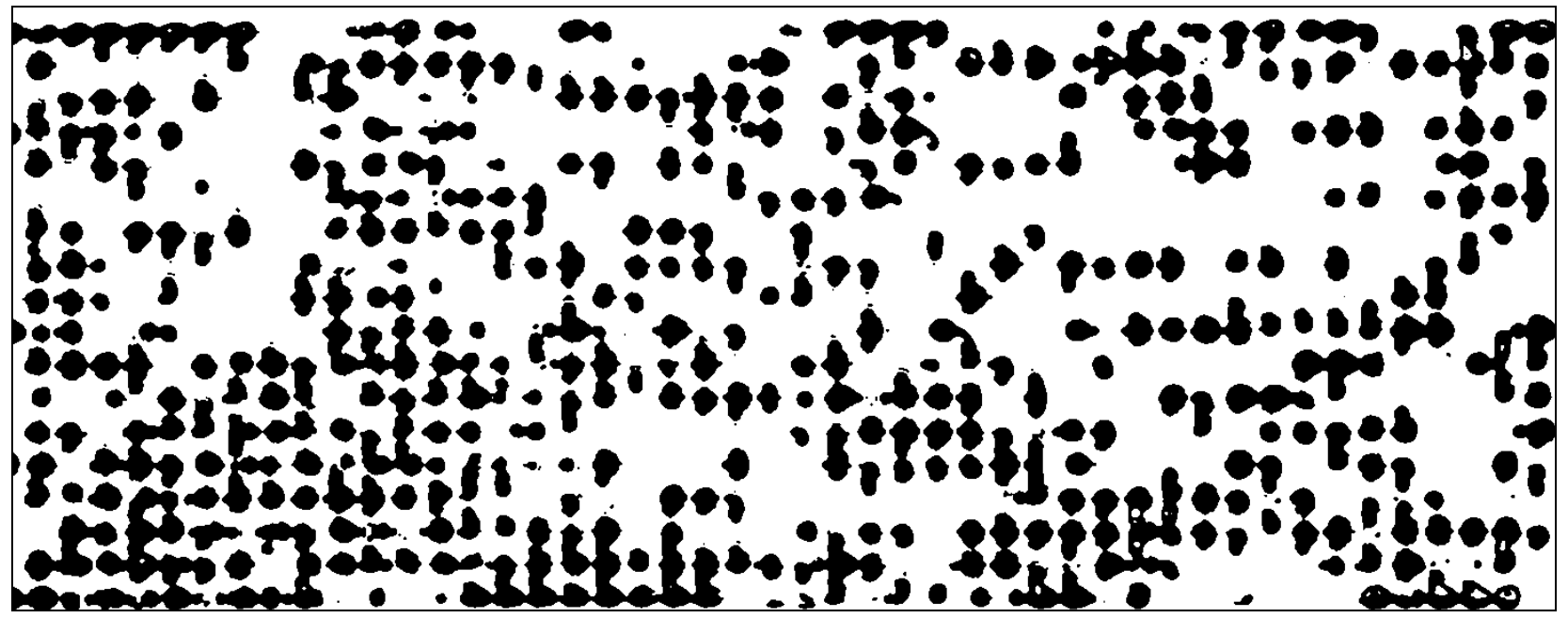}
\end{minipage}

\caption{\justifying
Representative example of the final image segmentation.
(a) Raw greyscale frame acquired under red back-illumination, where phase contrast arises from the refractive-index mismatch between silicone oil and air.
(b) Corresponding binarised image obtained after applying the trained LabKit classifier, with the air phase isolated.
}
\label{fig:segmentation}

\end{figure}

Recorded videos are segmented to isolate the spatial distribution of the air phase from the oil and solid phases. Because no dye is used, the contrast between the two phases is mainly located at fluid interfaces and arises from the refractive-index difference between the two fluids. Image segmentation is challenging because the air and oil phases have similar grey values, and image intensities along fluid interfaces are spatially variable. For this reason, we segment using the supervised machine-learning plugin LabKit in Fiji/ImageJ. A separate LabKit classifier is trained for each flow condition because changes in the phase distribution and contrast make a single classifier insufficient for the full dataset.
The procedure used to generate each classifier is illustrated in Fig.~\ref{fig:LabKit}.
For each flow condition, we first extract a representative image and define a region of interest for the analysis (Fig.~\ref{fig:LabKit}(a)).
Next, we manually label the representative image as foreground (air) and background (oil or solid) and train the classifier (Fig.~\ref{fig:LabKit}(b)).
Finally, the trained LabKit classifier is applied to the image stack to isolate the air phase as foreground (Fig.~\ref{fig:LabKit}(c)).
We apply the saved classifiers to the corresponding image stacks using the Python batch-processing script \texttt{Labkit\_batch\_segmentation.py}.
The full image-processing workflow is performed using \texttt{Complete\_Image\_to\_IS\_pipeline.ipynb}.
The saved LabKit classifiers and the batch-processing script are available on Zenodo~\citep{Dumay2026Zenodo}.
A representative example of the final segmentation result is shown in Fig.~\ref{fig:segmentation}.

\FloatBarrier

\subsection{Metrics of flow-reversal asymmetry}
\label{Spatial_Metric}

We introduce two measures for flow-reversal asymmetry, based on pressure
measurements and air-phase segmentation, respectively.
The pressure-based quantity $I_P$ compares the mean steady-state pressure drops
$\Delta P(Q)$ measured at the same imposed flow rate $Q$ in the AB and
BA directions:

\begin{equation}
I_P(Q) =
\frac{\Delta P_{BA}(Q)-\Delta P_{AB}(Q)}{P_0},
\end{equation}

where $P_0$ is the largest steady-state mean pressure drop measured across the full two-phase data set.
The same value of $P_0$ is used for T85, C85, and BL, allowing direct comparison between geometries.
Positive values of $I_P$ correspond to a larger pressure drop in the BA direction, whereas negative values correspond to a larger pressure drop in the AB direction.

A small directional mismatch remains at the lowest investigated flow rate for each geometry.
To account for this baseline mismatch, the lowest-flow value is used as an empirical reference, and we define

\begin{equation}
I_P^*(Q) = I_P(Q)-I_P(Q_{\min}),
\end{equation}

where $Q_{\min}$ is the lowest investigated flow rate for the corresponding geometry.

The repeatability contribution to the uncertainty in $I_P$ is estimated by independently bootstrapping the repetition-averaged pressure drops measured in the AB and BA directions.
The uncertainty in $I_P^*$ also includes the variability associated with the low-flow reference and the stated pressure-sensor accuracy ($\pm1.35\%$).
These contributions are treated as independent and combined in quadrature. 
Further details of the calculation are provided in the accompanying repository~\citep{Dumay2026Zenodo}.

\begin{figure}[htbp]
    \centering
    \includegraphics[width=\textwidth]
    {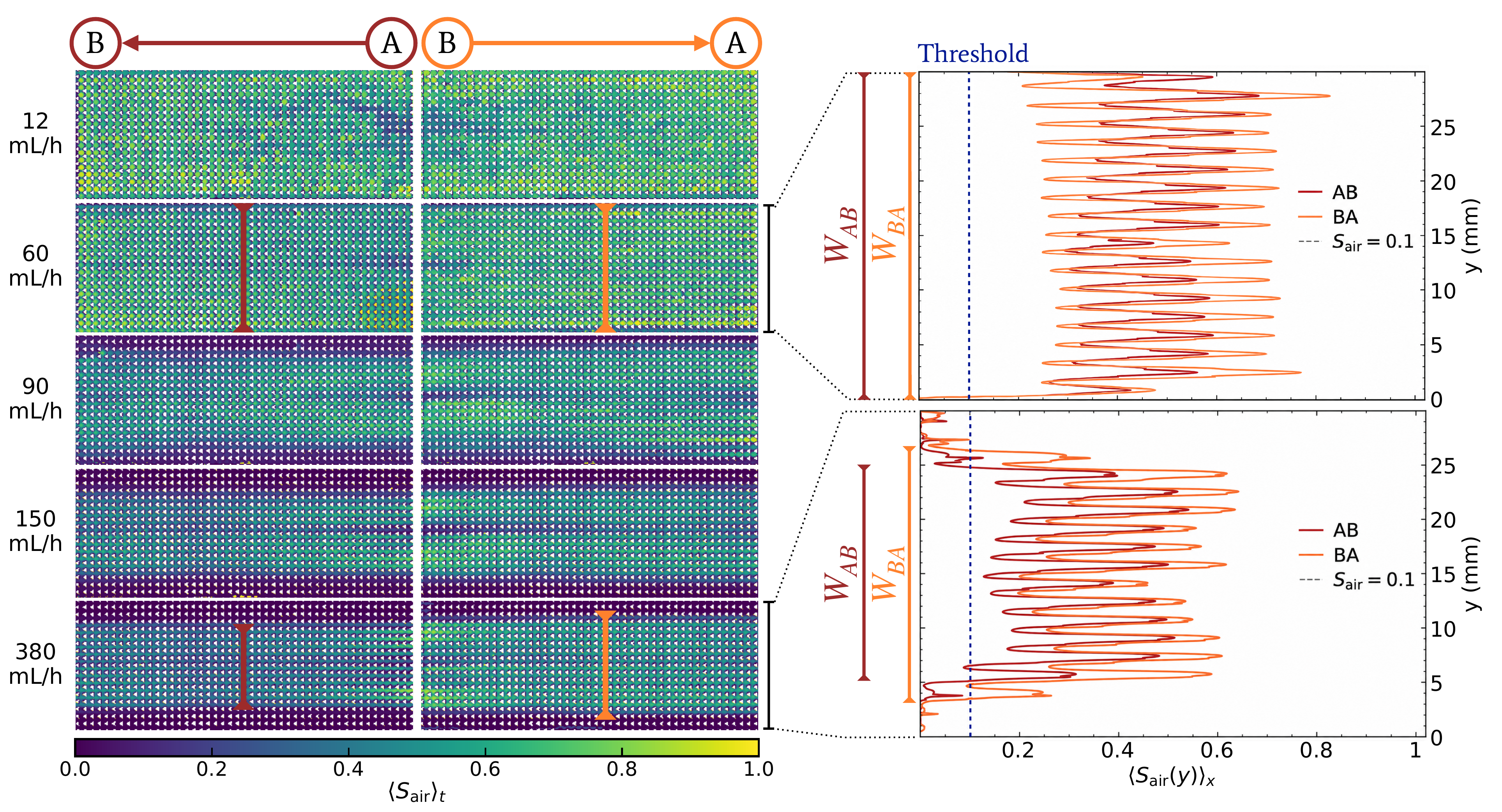}
    \caption{Extraction of active flow widths. Left: time-averaged air saturation maps for both flow directions at different flow rates. Right: transverse saturation profiles obtained by averaging along the flow direction. A common saturation threshold defines the active widths $W_{AB}$ and $W_{BA}$ used to calculate $I_S$.}
    \label{fig:satmaps}
\end{figure}

The image-based metric compares the active air-flow widths measured in the AB and BA directions.
Binary air-phase maps obtained by segmentation (Sec.~\ref{Image_analysis}) are averaged over the steady-state interval, so that each pixel value represents the probability of finding air at that location.
Averaging these maps along the flow direction gives transverse air-saturation profiles (Fig.~\ref{fig:satmaps}).

\begin{figure}[htbp]
    \centering
    \includegraphics[width=\textwidth]
    {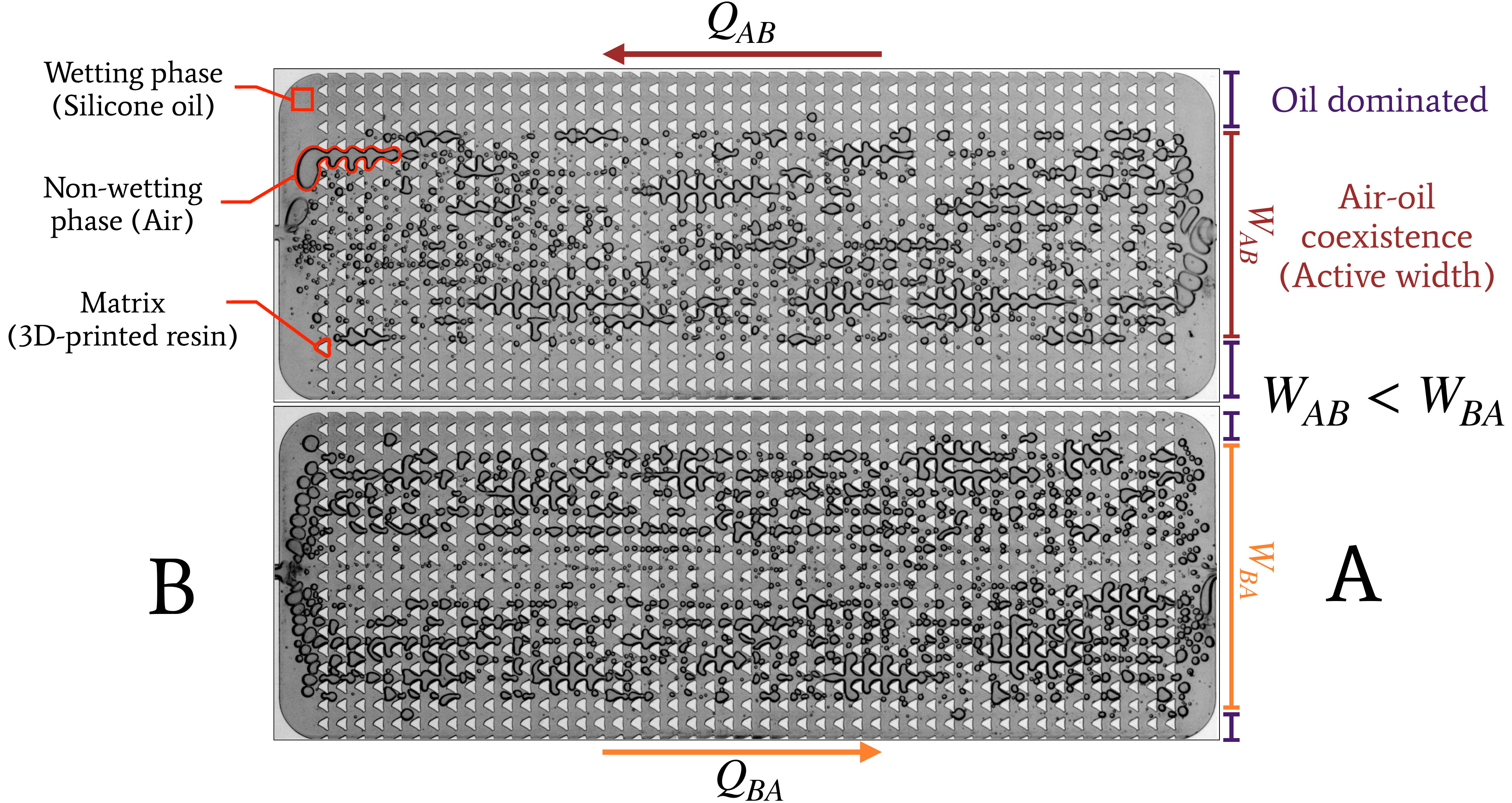}
    \caption{Active air-flow width $W$ for two-phase flow through triangular posts at $Q_{AB}=Q_{BA}=300\,\mathrm{mL\,h^{-1}}$. Flow reversal changes the active width, with $W_{AB}<W_{BA}$
    in the example shown.}
    \label{fig:Active_Width}
\end{figure}

For each repetition, the active width $W$ is defined as the lateral span between the first and last positions for which the time-averaged transverse air saturation exceeds a fixed threshold of $S_{\mathrm{th}}=0.10$.
The same threshold is applied to all flow rates, geometries, and directions (Fig.~\ref{fig:Active_Width}).

The image-based flow-reversal asymmetry is defined as

\begin{equation}
I_S(Q) =
\frac{W_{BA}(Q)-W_{AB}(Q)}
{W_{\mathrm{cell}}},
\end{equation}

where $W_{\mathrm{cell}}$ is the total analysed cell width. Thus, $I_S=0$ indicates identical active widths in the two directions.
Positive values indicate a larger active width in the BA direction, whereas negative values indicate a larger active width in the AB direction.

The uncertainty in $I_S$ accounts for experimental repeatability, sensitivity to the active-width threshold, and residual background asymmetry estimated from the low-flow BL and C85 measurements. These contributions are combined to obtain the error bars shown in the figures. Full details of the uncertainty calculation and data-processing procedure are provided in the accompanying Zenodo repository~\citep{Dumay2026Zenodo}. Pressure and image data are processed using \texttt{Plot\_Pressure\_Metric.ipynb} and \texttt{Complete\_Image\_to\_IS\_pipeline.ipynb}, respectively~\citep{Dumay2026Zenodo}.

\FloatBarrier
\subsection{Results}

\begin{figure}
\centering

\resizebox{\textwidth}{!}{%
\begin{tabular}{@{}c@{\hspace{0.15cm}}c@{}}

\includegraphics[height=5cm]
{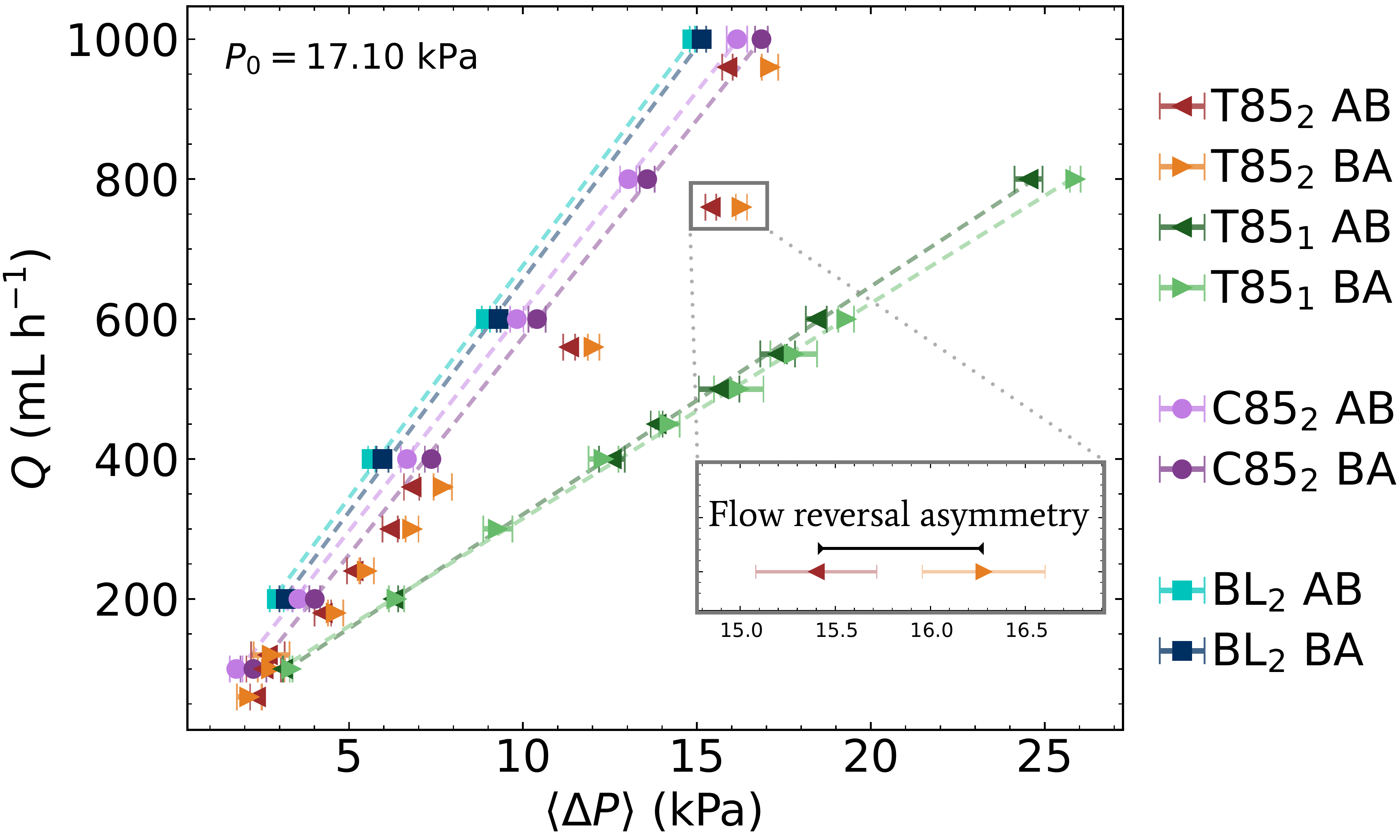}
&
\includegraphics[height=5cm]
{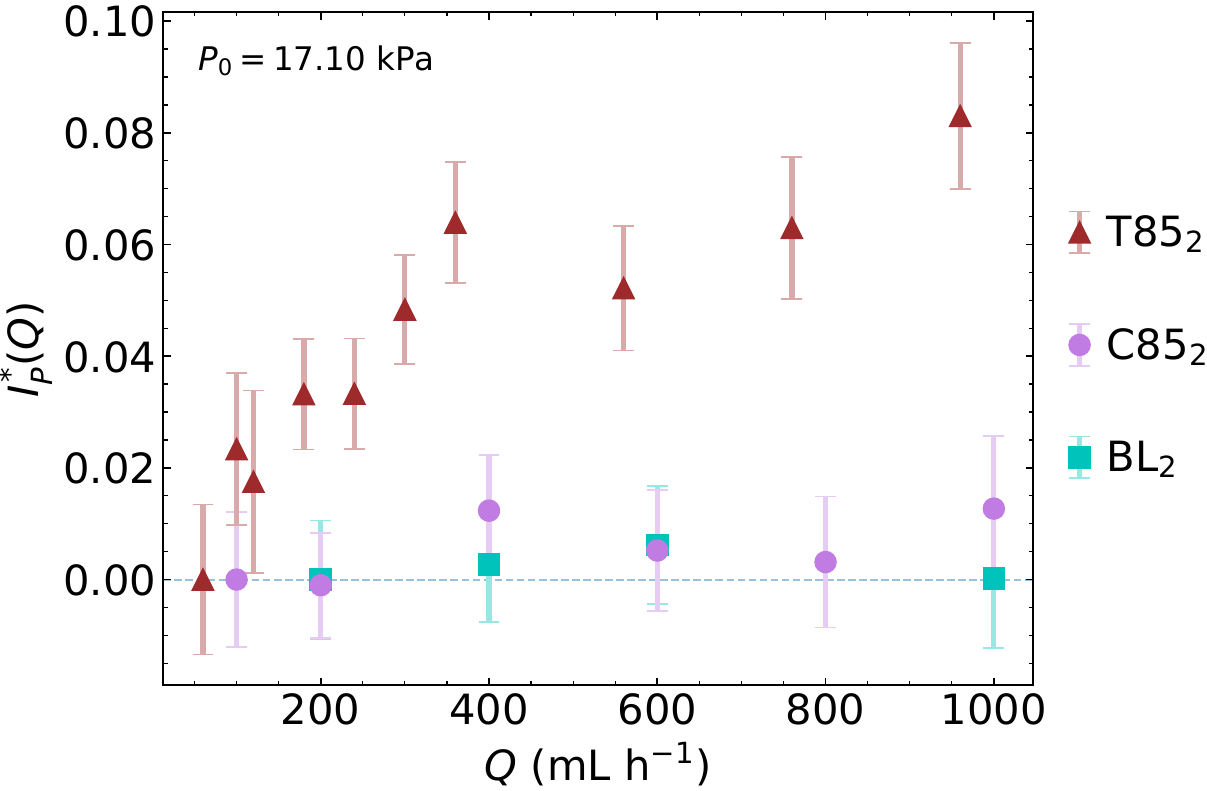}

\\[-0.05cm]

{\small (a)}
&
{\small (b)}

\end{tabular}%
}

\caption{\justifying
The numbers in the labels denote the number of flowing phases: 1 corresponds to single-phase oil flow, and 2 corresponds to two-phase oil--air flow.
(a) Total mean flow rate $Q$ as a function of the mean steady-state pressure drop $\Delta P$ for single-phase and two-phase flows measured in both directions.
(b) Corrected pressure-based flow-reversal asymmetry $I_P^*$ for the two-phase flows.
Structures C85 and BL, which have mirror-symmetric pore spaces, are flow-reversal symmetric within experimental uncertainty ($I_P^*(Q)\approx 0$). In contrast, T85 develops a positive flow-reversal asymmetry at higher flow rates ($Q_{\mathrm{tot}}\gtrsim 300\,\mathrm{mL\,h^{-1}}$).
}
\label{fig:press_combined}
\end{figure}

The pressure response is shown in
Fig.~\ref{fig:press_combined}.
Fig.~\ref{fig:press_combined}(a) reports the relation between the mean steady-state pressure drop and the imposed flow rate for the different cell geometries and flow configurations.

The single-phase pressure--flow curves remain linear over the investigated range, as expected in the creeping-flow regime.
The two-phase measurements for C85 and BL also remain approximately linear, consistent with the linear rheological regime reported for steady two-phase flow at sufficiently high capillary numbers~\citep{Berg2026,Hansen2026}.
The main exception is the T85 two-phase flow, which exhibits a pronounced non-linear pressure--flow relation at lower capillary numbers and approaches a more linear relation as the capillary number increases. Note that the absolute values of $\Delta P$ measured at the two highest $\mathrm{T85}_2$ flow rates in Fig.~\ref{fig:press_combined}(a) should be interpreted with caution because the zero offsets of the two pressure sensors were not correctly set for these measurements. This introduces a systematic offset in the absolute pressure-drop values. However, the same sensor settings were maintained for the corresponding AB and BA measurements, so the relative difference between the two flow directions remains reliable for assessing flow-reversal asymmetry. These points are therefore retained for the flow-reversal analysis, while their absolute pressure-drop values are not interpreted quantitatively.

Since panel (a) plots $Q$ versus $\Delta P$, a steeper curve corresponds to a lower apparent hydraulic resistance.
For a given geometry, introducing the non-wetting phase modifies the hydraulic response relative to the corresponding oil-only case, as expected from relative-permeability theory~\citep{bluntMultiphaseFlowPermeable2017}.
The smaller constrictions and asymmetric pore structure of T85 are expected to produce stronger capillary barriers and a broader distribution of local capillary-pressure thresholds.
As the forcing increases, progressive mobilisation of previously inactive flow paths may therefore contribute to the changes in slope observed for $\mathrm{T85}_2$, consistent with behaviour reported by~\citet{Fyhn2023}.

The differences between the AB and BA pressure responses remain relatively small when viewed directly in Fig.~\ref{fig:press_combined}(a), but become clearer when expressed through the corrected pressure-based flow-reversal asymmetry $I_P^*$ in Fig.~\ref{fig:press_combined}(b).
For T85, $I_P^*$ remains small at low flow rates and then increases with increasing flow rate, with a pronounced increase between $200\,\mathrm{mL\,h^{-1}}$ and $400\,\mathrm{mL\,h^{-1}}$. A small decrease is observed around $600\,\mathrm{mL\,h^{-1}}$; however, given the experimental uncertainty and limited sampling, this variation is not interpreted further. By comparison, $I_P^*$ remains small over the full investigated range for C85 and BL.

\begin{figure}
\centering

\resizebox{\textwidth}{!}{%
\begin{tabular}{@{}c@{\hspace{0.15cm}}c@{}}

\includegraphics[height=5cm]
{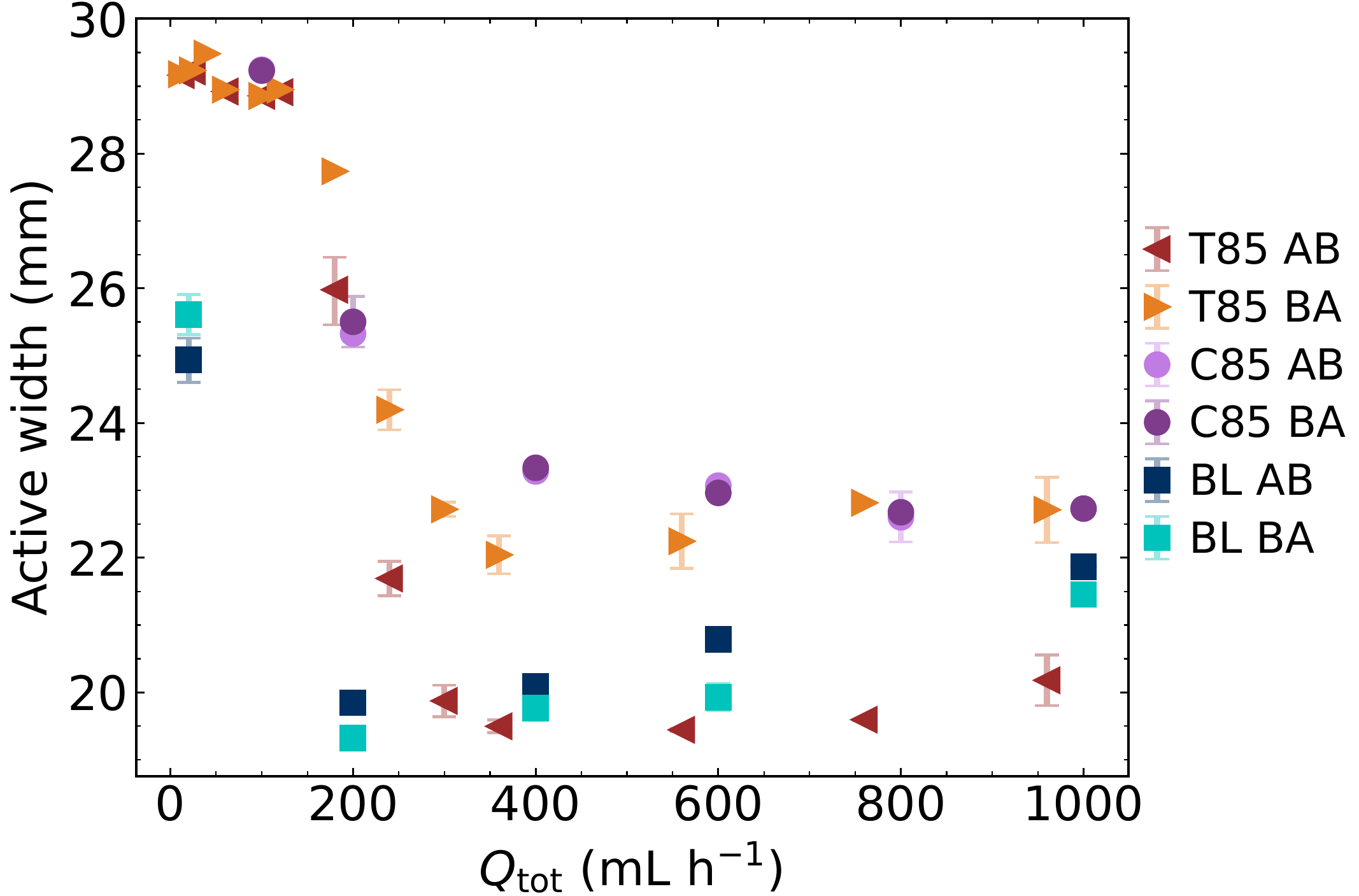}
&
\includegraphics[height=5cm]
{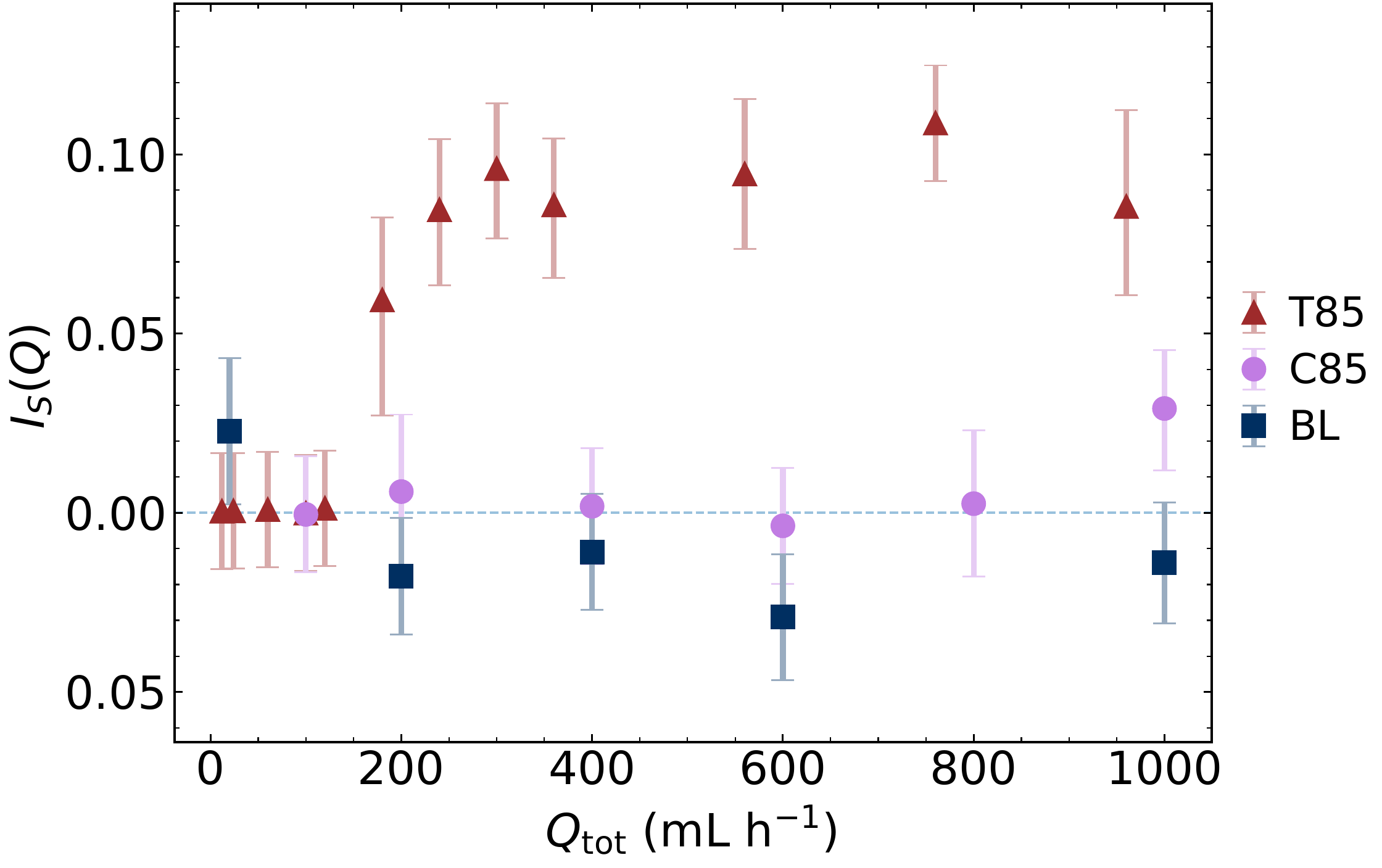}

\\[-0.05cm]

{\small (a)}
&
{\small (b)}

\end{tabular}%
}

\caption{\justifying
Image-based flow-reversal asymmetry response of the T85, C85, and BL cells.
(a) Active flow width as a function of flow rate for the AB and BA directions.
(b) Dimensionless image-based flow-reversal asymmetry $I_S$ as a function of flow rate. Here, C85 and BL are mainly symmetric within measurement uncertainty, while T85 shows asymmetry for $Q_{\mathrm{tot}}\gtrsim 300\,\mathrm{mL\,h^{-1}}$, mirroring the pressure-based conclusion in Fig.~\ref{fig:press_combined}(b).
}
\label{fig:width_combined}
\end{figure}

The image-based analysis shows a similar dependence on geometry.
Fig.~\ref{fig:width_combined}(a) shows the active flow widths in the AB and BA directions, while Fig.~\ref{fig:width_combined}(b) shows the corresponding image-based flow-reversal asymmetry $I_S$.

For T85, the active widths are similar in the two directions at low flow rates but progressively separate as the flow rate increases.
The BA direction exhibits a systematically larger active width than the AB direction over a substantial part of the investigated range.
In contrast, the AB and BA branches remain much closer for C85 and BL.

This behaviour is captured by $I_S$.
For T85, $I_S$ remains small at low flow rates and increases rapidly around $300\,\mathrm{mL\,h^{-1}}$, remaining elevated at higher flow rates.
For C85 and BL, $I_S$ remains close to zero over most of the investigated range.

Taken together, the pressure- and image-based metrics show the same overall geometry dependence: T85 exhibits a small but measurable flow-reversal asymmetry, whereas C85 and BL remain symmetric within the experimental uncertainty.
In T85, both the pressure and image-based metrics increase over a similar general flow-rate range, indicating that the directional pressure response is accompanied by a change in the spatial organisation of the two-phase flow.

\FloatBarrier

\section{Discussion}

\subsection{Fabrication protocol: capabilities and limitations}

The protocol presented here provides optical access qualitatively comparable to lithography-fabricated model porous media~\citep[e.g.,][]{Borgman2023}, while enabling rapid prototyping and customisation with consumer-grade equipment.
Because the pore geometry is defined digitally, arbitrary obstacle shapes, disorder levels, pore-size distributions, and inlet/outlet configurations could readily be implemented, such as thresholded Gaussian random fields~\citep{Hyman2014}, slices of natural porous materials~\citep{Buchgraber2012}, or other spatially heterogeneous structures~\citep{Bordoloi2022}.
The cells demonstrated here are quasi-two-dimensional with millimetre-scale pores, and surface treatment is applied to render the external surfaces optically transparent (Fig.~\ref{fig:Coating}).
All internal printed structures remain rough, which affects light transmission and is likely to modify local wetting properties~\citep{deGennes2004}.
In addition, the minimum practical pore size achievable with the present protocol is typically $\approx0.1\,\mathrm{mm}$.
Although consumer-grade stereolithography printers have stated resolutions to tens of micrometres~\citep{Au2016,Waheed2016}, over-curing into pores means this limit is rarely achievable in practice~\citep{macdonaldComparingMicrofluidicPerformance2017}.
Considerably smaller enclosed features have nevertheless been demonstrated using specialised stereolithography systems with custom optical configurations and resin formulations~\citep{Miner2026}.

Beyond internal roughness and resolution limits, the key fabrication bottlenecks for stereolithography-printed flow cells are the removal of uncured resin and the control of internal optical properties~\citep{Au2016,Waheed2016}.
Resin clearing becomes increasingly difficult in geometries with tortuous pathways or flow-stagnation zones, and this poses challenges to printing fully three-dimensional porous structures~\citep{Waheed2016,macdonaldComparingMicrofluidicPerformance2017}. Recent specialised DLP-SLA systems have nevertheless demonstrated enclosed three-dimensional structures with micrometre-scale pores~\citep{Miner2026}, showing that such geometries are technically achievable with dedicated printing hardware and resin formulations.
Direct optical imaging of such structures additionally requires matching the refractive indices of the solid structure
and working fluids~\citep{Haefeli2014}, complicated by the high refractive indices (typically $1.49$ or above) of stereolithography resins~\citep{Reynoso2021} and by spatial variability in the refractive indices of printed parts~\citep{Rank2019}.
Further efforts to produce monolithic flow cells with micrometre-scale or fully three-dimensional pore spaces would benefit from methods to better remove internal resin, reduce internal surface roughness, and minimise refractive-index inhomogeneities.

\subsection{Flow-reversal asymmetry in steady two-phase flow}

As a proof of concept, we studied steady simultaneous oil--air two-phase flow through cells containing triangular (T85), cylindrical (C85), and blank (BL) geometries fabricated with the protocol, and we quantified the flow-reversal asymmetry using both the macroscopic pressure response and the spatial organisation of the air phase.
The T85 geometry showed a clear directional dependence in both quantities, whereas C85 and BL remained symmetric within experimental uncertainty (see Fig.~\ref{fig:press_combined}~and~\ref{fig:width_combined}).
Because T85 and C85 have the same porosity but different obstacle shape and pore geometry, these results support the conclusion that additional pore-space characteristics besides porosity are required to determine the constitutive relation for steady two-phase flows~\citep{Armstrong2016,Valavanides2018,Berg2026}.

In single-phase Newtonian creeping flow, the hydraulic response is reciprocal under flow reversal, while direction-dependent hydraulic resistance in asymmetric conduits and architected porous media generally emerges when inertial effects become relevant~\citep{Nguyen2021,Bohm2022,Vezies2026}.
In two-phase flow, the constitutive relation between flow rate and pressure drop can be either linear or non-linear depending on the capillary number~\citep{TallakstadPRL2009,SinhaEtAl2017}.
Ultimately, this rheology originates from the microstructural arrangement of fluid interfaces in the pore space, which controls capillary-pressure thresholds, trapping, and changes in phase connectivity, even in low Reynolds number flows~\citep{Hansen2026,Berg2026}.
Our results suggest that asymmetric pore geometries can couple to these interfacial processes strongly enough to produce a different hydraulic response when the direction of the imposed flow is reversed, likely because the separation distances and orientations of the pore walls establish different capillary-entry pressures for advancing fluid interfaces in each flow direction.
A similar directional dependence of capillary thresholds is familiar from studies of drainage--imbibition hysteresis~\citep{JoekarNiasar2013,Schluter2016}.

Consistent with this picture, the triangular geometry (T85) departs most strongly from a Darcy-like linear pressure--flow relation~\citep{SinhaHansen2012,SinhaEtAl2017}, suggesting that the progressive mobilisation of interfaces and flow pathways may contribute to the directional response. Future studies should test whether the degree of flow-reversal asymmetry correlates with changes in the effective two-phase rheology~\citep{Berg2026,Hansen2026}. 
The present proof-of-concept experiments are limited to two porous geometries at the same designed porosity, one fluid pair, one wettability condition, and an equal phase-flow ratio. They therefore do not establish which particular aspects of the triangular pore geometry are responsible for the observed asymmetry, nor how general the effect is across different two-phase systems. Future studies could independently vary obstacle shape and orientation, pore-size distribution, disorder, fractional flow, viscosity ratio, wettability, and capillary number to identify the geometrical and fluid parameters controlling the onset and magnitude of flow-reversal asymmetry. More detailed image analysis could additionally quantify local saturation, phase connectivity, interface length, and pore-by-pore occupancy, allowing changes in the macroscopic pressure response to be related more directly to rearrangements of the two-phase configuration.

\section{Conclusion}

We have presented an open-source protocol combining stereolithography 3D printing, internal flushing, and external resin coating to produce transparent millifluidic flow cells with easily customisable geometries.
As a proof of concept, we used the cells to study steady oil--air two-phase flow through porous geometries with different obstacle shapes.
Both pressure- and image-based measurements showed that the triangular T85 geometry exhibits a clear directional dependence under flow reversal, whereas the cylindrical C85 and blank BL geometries remain symmetric within experimental uncertainty. 
The observed breaking of flow-reversal symmetry accompanies the broken mirror symmetry of the triangular obstacles at the pore scale, likely because the separation distances and orientations of pore walls control the microstructural arrangement of fluid interfaces.
Designing pore structures that enhance this flow asymmetry is an attractive prospect for future studies.
Beyond its fundamental interest, flow-reversal asymmetry could potentially be optimised through pore-space design and exploited for passive two-phase flow control. For example, sufficiently strong directional differences in hydraulic resistance could provide the basis for a two-phase fluidic diode, analogous in function to Tesla-type valves~\citep{Nguyen2021,Bohm2022}. Unlike such single-phase devices, where direction-dependent resistance generally relies on inertial effects, the mechanism considered here could operate at low Reynolds numbers through capillary and interfacial effects. Because the directional response arises from pore-scale geometry and interfacial dynamics, this approach could potentially enable compact passive flow-control elements.

The protocol is intended to lower the barrier to fabricating transparent millifluidic flow cells, while the proof-of-concept experiments demonstrate that pore-scale geometry alone can introduce a directional dependence into the constitutive relation between pressure drop and flow rate in steady two-phase flow.
All design files, processing scripts, and analysis software are openly available in the accompanying Zenodo repository~\citep{Dumay2026Zenodo}, providing an accessible platform for other laboratories to reproduce these cells and adapt them to diverse porous media studies.

\section*{Acknowledgements}
The authors thank Lydéric Bocquet, Marie-Caroline Jullien, Daan Frenkel, and Marcel Moura for many stimulating discussions. The authors are also grateful to Daan Frenkel, Paula Reis, and Andreas Andersen Hennig for carefully reading the manuscript and providing constructive comments, to Pooja Singh for support with data acquisition, and to Vitor Heitor Cunha for his feedback on the graphical abstract.

\bibliography{refs}

@article{karadimitriou2012review,
  title={A review of micromodels and their use in two-phase flow studies},
  author={Karadimitriou, NK and Hassanizadeh, SM},
  journal={Vadose Zone Journal},
  volume={11},
  number={3},
  pages={vzj2011.0072},
  year={2012},
  publisher={The Soil Science Society of America, Inc.},
  doi       = {10.2136/vzj2011.0072}
}

@article{qin2010soft,
  title={Soft lithography for micro- and nanoscale patterning},
  author={Qin, Dong and Xia, Younan and Whitesides, George M},
  journal={Nature protocols},
  volume={5},
  number={3},
  pages={491–502},
  year={2010},
  publisher={Nature Publishing Group},
  doi       = {10.1038/nprot.2009.234}
}

@article{jahanbakhsh2020review,
  title={Review of microfluidic devices and imaging techniques for fluid flow study in porous geomaterials},
  author={Jahanbakhsh, Amir and Wlodarczyk, Krystian L and Hand, Duncan P and Maier, Robert RJ and Maroto-Valer, M Mercedes},
  journal={Sensors},
  volume={20},
  number={14},
  pages={4030},
  year={2020},
  publisher={MDPI},
  doi       = {10.3390/s20144030}
}

@article{PatsoukisDimou2022,
  author  = {Patsoukis Dimou, Alexandros and Menke, Hannah P. and Maes, Julien},
  title   = {Benchmarking the Viability of 3D Printed Micromodels for Single Phase Flow Using Particle Image Velocimetry and Direct Numerical Simulations},
  journal = {Transport in Porous Media},
  volume  = {141},
  pages   = {279--294},
  year    = {2022},
  doi     = {10.1007/s11242-021-01718-8}
}

@article{raeini2014direct,
  title={Direct simulations of two-phase flow on micro-CT images of porous media and upscaling of pore-scale forces},
  author={Raeini, Ali Q and Blunt, Martin J and Bijeljic, Branko},
  journal={Advances in water resources},
  volume={74},
  pages={116--126},
  year={2014},
  publisher={Elsevier},
  doi       = {10.1016/j.advwatres.2014.08.012}
}

@article{silin2011microtomography,
  title={Microtomography and pore-scale modeling of two-phase fluid distribution},
  author={Silin, Dmitriy and Tomutsa, Liviu and Benson, Sally M and Patzek, Tad W},
  journal={Transport in porous media},
  volume={86},
  number={2},
  pages={495--515},
  year={2011},
  publisher={Springer},
  doi       = {10.1007/s11242-010-9636-2}
}

@article{tekseth20244d,
  title={4D imaging of two-phase flow in porous media using laboratory-based micro-computed tomography},
  author={Tekseth, KR and Breiby, DW},
  journal={Water Resources Research},
  volume={60},
  number={4},
  pages={e2023WR036514},
  year={2024},
  publisher={Wiley Online Library},
  doi       = {10.1029/2023WR036514}
}

@book{deGennes2004,
  author    = {de Gennes, Pierre-Gilles and Brochard-Wyart, Fran{\c{c}}oise and Qu{\'e}r{\'e}, David},
  title     = {Capillarity and Wetting Phenomena: Drops, Bubbles, Pearls, Waves},
  publisher = {Springer},
  address   = {New York},
  year      = {2004},
  doi       = {10.1007/978-0-387-21656-0}
}

@book{Feder2022,
  author    = {Feder, Jens and Flekk{\o}y, Eirik G. and Hansen, Alex},
  title     = {Physics of Flow in Porous Media},
  publisher = {Cambridge University Press},
  address   = {Cambridge},
  year      = {2022},
  doi       = {10.1017/9781009100717}
}

@book{Bear1972,
author    = {Bear, Jacob},
title     = {Dynamics of Fluids in Porous Media},
publisher = {Elsevier},
year      = {1972}
}

@article{Fyhn2023,
author  = {Fyhn, Hursanay and Sinha, Santanu and Hansen, Alex},
title   = {Effective Rheology of Immiscible Two-Phase Flow in Porous Media Consisting of Random Mixtures of Grains Having Two Types of Wetting Properties},
journal = {Frontiers in Physics},
volume  = {11},
pages   = {1175426},
year    = {2023},
doi     = {10.3389/fphy.2023.1175426}
}

@article{TallakstadPRL2009,
author  = {Tallakstad, Ken Tore and Knudsen, Henning Arendt and Ramstad, Thomas and L{\o}voll, Grunde and M{\aa}l{\o}y, Knut J{\o}rgen and Toussaint, Renaud and Flekk{\o}y, Eirik Grude},
title   = {Steady-State Two-Phase Flow in Porous Media: Statistics and Transport Properties},
journal = {Physical Review Letters},
volume  = {102},
number  = {7},
pages   = {074502},
year    = {2009},
doi     = {10.1103/PhysRevLett.102.074502}
}

@article{TallakstadPRE2009,
author  = {Tallakstad, Ken Tore and L{\o}voll, Grunde and Knudsen, Henning Arendt and Ramstad, Thomas and Flekk{\o}y, Eirik Grude and M{\aa}l{\o}y, Knut J{\o}rgen},
title   = {Steady-State, Simultaneous Two-Phase Flow in Porous Media: An Experimental Study},
journal = {Physical Review E},
volume  = {80},
number  = {3},
pages   = {036308},
year    = {2009},
doi     = {10.1103/PhysRevE.80.036308}
}

@article{Gerami2019,
  author  = {Gerami, Alireza and Alzahid, Yara and Mostaghimi, Peyman and Kashaninejad, Navid and Kazemifar, Farzan and Amirian, Tammy and Mosavat, Nader and Ebrahimi Warkiani, Majid and Armstrong, Ryan T.},
  title   = {Microfluidics for Porous Systems: Fabrication, Microscopy and Applications},
  journal = {Transport in Porous Media},
  volume  = {130},
  pages   = {277--304},
  year    = {2019},
  doi     = {10.1007/s11242-018-1202-3}
}

@article{Valavanides2018,
author  = {Valavanides, M. S.},
title   = {Review of Steady-State Two-Phase Flow in Porous Media: Independent Variables, Universal Energy Efficiency Map, Critical Flow Conditions, Effective Characterization of Flow and Pore Network},
journal = {Transport in Porous Media},
volume  = {123},
number  = {1},
pages   = {45--99},
year    = {2018},
doi     = {10.1007/s11242-018-1026-1}
}

@article{JoekarNiasar2013,
  author = {Joekar-Niasar, Vahid and Doster, Florian and Armstrong, Ryan T. and Wildenschild, Dorthe and Celia, Michael A.},
title   = {Trapping and Hysteresis in Two-Phase Flow in Porous Media: A Pore-Network Study},
journal = {Water Resources Research},
volume  = {49},
number  = {7},
pages   = {4244--4256},
year    = {2013},
doi     = {10.1002/wrcr.20313}
}

@article{Schluter2016,
author  = {Schlüter, Steffen and Berg, Steffen and Rücker, Matthias and Armstrong, Ryan T. and Vogel, Hans-Jörg and Hilfer, Rolf and Wildenschild, Dorthe},
title   = {Pore-Scale Displacement Mechanisms as a Source of Hysteresis for Two-Phase Flow in Porous Media},
journal = {Water Resources Research},
volume  = {52},
number  = {3},
pages   = {2194--2205},
year    = {2016},
doi     = {10.1002/2015WR018254}
}

@article{SinhaHansen2012,
author  = {Sinha, Santanu and Hansen, Alex},
title   = {Effective Rheology of Immiscible Two-Phase Flow in Porous Media},
journal = {Europhysics Letters},
volume  = {99},
number  = {4},
pages   = {44004},
year    = {2012},
doi     = {10.1209/0295-5075/99/44004}
}

@article{SinhaEtAl2017,
  author = {Sinha, Santanu and Bender, Andrew T. and Danczyk, Matthew and Keepseagle, Kayla and Prather, Cody A. and Bray, Joshua M. and Thrane, Linn W. and Seymour, Joseph D. and Codd, Sarah L. and Hansen, Alex},
title   = {Effective Rheology of Two-Phase Flow in Three-Dimensional Porous Media: Experiment and Simulation},
journal = {Transport in Porous Media},
volume  = {119},
number  = {1},
pages   = {77--94},
year    = {2017},
doi     = {10.1007/s11242-017-0874-4}
}

@article{Sales2022,
author  = {Sales, J. M. A. and Seybold, H. J. and Oliveira, C. L. N. and Andrade, J. S.},
title   = {Bubble Dynamics in Stationary Two-Phase Flow Through Disordered Porous Media},
journal = {Frontiers in Physics},
volume  = {10},
pages   = {860190},
year    = {2022},
doi     = {10.3389/fphy.2022.860190}
}

@article{Nielsen2020,
author  = {Nielsen, Anna V. and Beauchamp, Michael J. and Nordin, Gregory P. and Woolley, Adam T.},
title   = {3D Printed Microfluidics},
journal = {Annual Review of Analytical Chemistry},
volume  = {13},
pages   = {45--65},
year    = {2020},
doi     = {10.1146/annurev-anchem-091619-102649}
}

@article{Musgrove2022,
author  = {Musgrove, Hannah B. and Catterton, Megan A. and Pompano, Rebecca R.},
title   = {Applied Tutorial for the Design and Fabrication of Biomicrofluidic Devices by Resin 3D Printing},
journal = {Analytica Chimica Acta},
volume  = {1209},
pages   = {339842},
year    = {2022},
doi     = {10.1016/j.aca.2022.339842}
}

@article{Kecili2020,
author  = {Ke{\c{c}}ili, Seren and Tekin, H. Cumhur},
title   = {Adhesive Bonding Strategies to Fabricate High-Strength and Transparent 3D Printed Microfluidic Device},
journal = {Biomicrofluidics},
volume  = {14},
number  = {2},
pages   = {024113},
year    = {2020},
doi     = {10.1063/5.0003302}
}

@article{VahidDastjerdi2022,
  author = {Vahid Dastjerdi, Samaneh and Karadimitriou, Nikolaos and Hassanizadeh, S. Majid and Steeb, Holger},
title   = {Experimental Evaluation of Fluid Connectivity in Two-Phase Flow in Porous Media During Drainage},
journal = {Water Resources Research},
volume  = {58},
number  = {11},
pages   = {e2022WR033451},
year    = {2022},
doi     = {10.1029/2022WR033451}
}

@article{JoekarNiasar2012,
author  = {Joekar-Niasar, Vahid and Hassanizadeh, S. Majid},
title   = {Analysis of Fundamentals of Two-Phase Flow in Porous Media Using Dynamic Pore-Network Models: A Review},
journal = {Critical Reviews in Environmental Science and Technology},
volume  = {42},
number  = {18},
pages   = {1895--1976},
year    = {2012},
doi     = {10.1080/10643389.2011.574101}
}

@article{Khobaib2025,
author  = {Khobaib, Khobaib and Reis, Paula and Moura, Marcel and Toussaint, Renaud and Flekk{\o}y, Eirik Grude and M{\aa}l{\o}y, Knut J{\o}rgen},
title   = {Gravity stabilized drainage in porous media with controlled disorder},
journal = {Physical Review Research},
volume  = {7},
pages   = {023040},
year    = {2025},
doi     = {10.1103/PhysRevResearch.7.023040}
}

@article{Au2016,
author  = {Au, Anthony K. and Huynh, Wilson and Lisa F. Horowitz. and Folch, Albert},
title   = {3D-printed microfluidics},
journal = {Angewandte Chemie International Edition},
volume  = {55},
number  = {12},
pages   = {3862--3881},
year    = {2016},
doi     = {10.1002/anie.201504382}
}

@article{Bohm2022,
  author  = {Bohm, Sebastian and Phi, Hai Binh and Moriyama, Ayaka and Runge, Erich and Strehle, Steffen and K{\"o}nig, J{\"o}rg and Cierpka, Christian and Dittrich, Lars},
  title   = {Highly efficient passive Tesla valves for microfluidic applications},
  journal = {Microsystems \& Nanoengineering},
  volume  = {8},
  pages   = {97},
  year    = {2022},
  doi     = {10.1038/s41378-022-00437-4}
}

@article{Nguyen2021,
author  = {Nguyen, Quynh M. and Abouezzi, Joanna and Ristroph, Leif},
title   = {Early turbulence and pulsatile flows enhance diodicity of Tesla's macrofluidic valve},
journal = {Nature Communications},
volume  = {12},
pages   = {2884},
year    = {2021},
doi     = {10.1038/s41467-021-23009-y}
}

@article{Waheed2016,
  author = {Waheed, Sidra and Cabot, Joan M. and Macdonald, Niall P. and Lewis, Trevor and Guijt, Rosanne M. and Paull, Brett and Breadmore, Michael C.},
title   = {3D printed microfluidic devices: enablers and barriers},
journal = {Lab on a Chip},
volume  = {16},
pages   = {1993--2013},
year    = {2016},
doi     = {10.1039/C6LC00284F}
}

@misc{Dumay2026Zenodo,
  author    = {Dumay, Yann and Pierce, J. Kevin and Eiser, Erika},
  title     = {Transparent 3D-printed flow-cell designs, experimental data,
               and analysis code for two-phase flow-reversal experiments},
  year      = {2026},
  publisher = {Zenodo},
  doi       = {10.5281/zenodo.21836007},
}

@article{Maier2008,
  author = {Maier, Robert S. and Schure, Mark R. and Gage, Justin P. and Seymour, Joseph D.},
  title   = {Sensitivity of pore-scale dispersion to the construction of random bead packs},
  journal = {Water Resources Research},
  volume  = {44},
  number  = {6},
  pages   = {W06S03},
  year    = {2008},
  doi     = {10.1029/2006WR005577}
}

@article{souzy2020,
  author  = {Souzy, Mathieu and Lhuissier, Henri and M{\'e}heust, Yves and Le Borgne, Tanguy and Metzger, Bloen},
  title   = {Velocity distributions, dispersion and stretching in three-dimensional porous media},
  journal = {Journal of Fluid Mechanics},
  volume  = {891},
  pages   = {A16},
  year    = {2020},
  doi     = {10.1017/jfm.2020.113}
}

@article{heyman2021,
  author  = {Heyman, Joris and Lester, Daniel R. and Le Borgne, Tanguy},
  title   = {Scalar Signatures of Chaotic Mixing in Porous Media},
  journal = {Physical Review Letters},
  volume  = {126},
  number  = {3},
  pages   = {034505},
  year    = {2021},
  doi     = {10.1103/PhysRevLett.126.034505}
}

@article{brodin2022,
  author  = {Brodin, Joachim Falck and Rikvold, Per Arne and Moura, Marcel and Toussaint, Renaud and M{\aa}l{\o}y, Knut J{\o}rgen},
  title   = {Competing Gravitational and Viscous Effects in 3D Two-Phase Flow Investigated With a Table-Top Optical Scanner},
  journal = {Frontiers in Physics},
  volume  = {10},
  pages   = {936915},
  year    = {2022},
  doi     = {10.3389/fphy.2022.936915}
}

@article{Krevor2023,
  author  = {Krevor, Samuel and de Coninck, Heleen and Gasda, Sarah E. and Ghaleigh, Navraj Singh and de Gooyert, Vincent and Hajibeygi, Hadi and Juanes, Ruben and Neufeld, Jerome and Roberts, Jennifer J. and Swennenhuis, Floris},
  title   = {Subsurface carbon dioxide and hydrogen storage for a sustainable energy future},
  journal = {Nature Reviews Earth \& Environment},
  volume  = {4},
  pages   = {102--118},
  year    = {2023},
  doi     = {10.1038/s43017-022-00376-8}
}

@article{LeBorgneHeyman2026,
  author  = {Le Borgne, Tanguy and Heyman, Joris},
  title   = {Fluid Deformation and Mixing in Porous Media as Drivers for Chemical and Biological Processes},
  journal = {Annual Review of Fluid Mechanics},
  volume  = {58},
  pages   = {443--471},
  year    = {2026},
  doi     = {10.1146/annurev-fluid-112723-051940}
}

@article{Heinemann2021,
  author  = {Heinemann, Niklas and Alcalde, Juan and Miocic, Johannes M. and Hangx, Suzanne J. T. and Kallmeyer, Jens and Ostertag-Henning, Christian and Hassanpouryouzband, Aliakbar and Thaysen, Eike M. and Strobel, Gion J. and Schmidt-Hattenberger, Cornelia and Edlmann, Katriona and Wilkinson, Mark and Bentham, Michelle and Haszeldine, R. Stuart and Carbonell, Ramon and Rudloff, Alexander},
  title   = {Enabling large-scale hydrogen storage in porous media -- the scientific challenges},
  journal = {Energy \& Environmental Science},
  volume  = {14},
  pages   = {853--864},
  year    = {2021},
  doi     = {10.1039/D0EE03536J}
}

@article{ZamelLi2013,
  author  = {Zamel, Nada and Li, Xianguo},
  title   = {Effective transport properties for polymer electrolyte membrane fuel cells -- With a focus on the gas diffusion layer},
  journal = {Progress in Energy and Combustion Science},
  volume  = {39},
  number  = {1},
  pages   = {111--146},
  year    = {2013},
  doi     = {10.1016/j.pecs.2012.07.002}
}

@article{Mukherjee2011,
  author  = {Mukherjee, Partha P. and Kang, Qinjun and Wang, Chao-Yang},
  title   = {Pore-scale modeling of two-phase transport in polymer electrolyte fuel cells -- progress and perspective},
  journal = {Energy \& Environmental Science},
  volume  = {4},
  number  = {2},
  pages   = {346--369},
  year    = {2011},
  doi     = {10.1039/B926077C}
}

@article{JinSengupta2024,
  author  = {Jin, Chenyu and Sengupta, Anupam},
  title   = {Microbes in porous environments: from active interactions to emergent feedback},
  journal = {Biophysical Reviews},
  volume  = {16},
  pages   = {173--188},
  year    = {2024},
  doi     = {10.1007/s12551-024-01185-7}
}

@article{TeconOr2017,
  author  = {Tecon, Robin and Or, Dani},
  title   = {Biophysical processes supporting the diversity of microbial life in soil},
  journal = {FEMS Microbiology Reviews},
  volume  = {41},
  number  = {5},
  pages   = {599--623},
  year    = {2017},
  doi     = {10.1093/femsre/fux039}
}

@article{Borgman2023,
  author  = {Borgman, Oshri and Turuban, R{\'e}gis and G{\'e}raud, Baudouin and Le Borgne, Tanguy and M{\'e}heust, Yves},
  title   = {Solute Front Shear and Coalescence Control Concentration Gradient Dynamics in Porous Micromodel},
  journal = {Geophysical Research Letters},
  volume  = {50},
  number  = {5},
  pages   = {e2022GL101407},
  year    = {2023},
  doi     = {10.1029/2022GL101407}
}

@article{Lovoll2004,
  author  = {L{\o}voll, Grunde and M{\'e}heust, Yves and Toussaint, Renaud and Schmittbuhl, Jean and M{\aa}l{\o}y, Knut J{\o}rgen},
  title   = {Growth activity during fingering in a porous Hele-Shaw cell},
  journal = {Physical Review E},
  volume  = {70},
  number  = {2},
  pages   = {026301},
  year    = {2004},
  doi     = {10.1103/PhysRevE.70.026301}
}

@article{Wright2017,
  author  = {Wright, Stuart F. and Zadrazil, Ivan and Markides, Christos N.},
  title   = {A review of solid--fluid selection options for optical-based measurements in single-phase liquid, two-phase liquid--liquid and multiphase solid--liquid flows},
  journal = {Experiments in Fluids},
  volume  = {58},
  pages   = {108},
  year    = {2017},
  doi     = {10.1007/s00348-017-2386-y}
}

@article{Datta2014,
  author  = {Datta, Sujit S. and Dupin, Jean-Baptiste and Weitz, David A.},
  title   = {Fluid breakup during simultaneous two-phase flow through a three-dimensional porous medium},
  journal = {Physics of Fluids},
  volume  = {26},
  number  = {6},
  pages   = {062004},
  year    = {2014},
  doi     = {10.1063/1.4884955}
}

@article{DoNascimento2019,
  author  = {do Nascimento, D{\'e}bora F. and Vimieiro Junior, Jos{\'e} R. and Paciornik, Sidnei and Carvalho, Marcio S.},
  title   = {Pore Scale Visualization of Drainage in 3D Porous Media by Confocal Microscopy},
  journal = {Scientific Reports},
  volume  = {9},
  pages   = {12333},
  year    = {2019},
  doi     = {10.1038/s41598-019-48803-z}
}

@article{Kim2023,
  author  = {Kim, Yong Tae and Ahmadianyazdi, Alireza and Folch, Albert},
  title   = {A print--pause--print protocol for 3D printing microfluidics using multimaterial stereolithography},
  journal = {Nature Protocols},
  volume  = {18},
  pages   = {1243--1259},
  year    = {2023},
  doi     = {10.1038/s41596-022-00792-6}
}

@article{VincentDospital2022,
  author  = {Vincent-Dospital, Tom and Moura, Marcel and Toussaint, Renaud and M{\aa}l{\o}y, Knut J{\o}rgen},
  title   = {Stable and unstable capillary fingering in porous media with a gradient in grain size},
  journal = {Communications Physics},
  volume  = {5},
  pages   = {306},
  year    = {2022},
  doi     = {10.1038/s42005-022-01072-1}
}

@article{Zhang2025,
  author  = {Zhang, Xueyi and Dou, Zhi and Hamada, Mayumi and de Anna, Pietro and Jimenez-Martinez, Joaquin},
  title   = {Enhanced Reaction Kinetics in Stationary Two-Phase Flow through Porous Media},
  journal = {Environmental Science \& Technology},
  volume  = {59},
  number  = {2},
  pages   = {1334--1343},
  year    = {2025},
  doi     = {10.1021/acs.est.4c09449}
}

@article{Maloy2021,
  author  = {M{\aa}l{\o}y, Knut J{\o}rgen and Moura, Marcel and Hansen, Alex and Flekk{\o}y, Eirik Grude and Toussaint, Renaud},
  title   = {Burst Dynamics, Upscaling and Dissipation of Slow Drainage in Porous Media},
  journal = {Frontiers in Physics},
  volume  = {9},
  pages   = {796019},
  year    = {2021},
  doi     = {10.3389/fphy.2021.796019}
}

@article{Anbari2018,
  author  = {Anbari, Alimohammad and Chien, Hung-Ta and Datta, Sujit S. and Deng, Wen and Weitz, David A. and Fan, Jing},
  title   = {Microfluidic Model Porous Media: Fabrication and Applications},
  journal = {Small},
  volume  = {14},
  number  = {18},
  pages   = {1703575},
  year    = {2018},
  doi     = {10.1002/smll.201703575}
}

@article{Hansen2026,
  author  = {Hansen, Alex and Sinha, Santanu},
  title   = {Immiscible Two-Phase Flow in Porous Media: A Statistical Mechanics Approach},
  journal = {arXiv preprint arXiv:2603.09658},
  year    = {2026},
  doi     = {10.48550/arXiv.2603.09658}
}

@article{Pierce2026,
  author  = {Pierce, J. Kevin and Le Borgne, Tanguy and Renard, Fran{\c{c}}ois and Linga, Gaute},
  title   = {How Pore-Scale Disorder Controls Fluid Stretching in Porous Media},
  journal = {arXiv preprint arXiv:2604.02981},
  year    = {2026},
  doi     = {10.48550/arXiv.2604.02981}
}

@article{Linga2026,
  author  = {Linga, Gaute and Pierce, Kevin and Moura, Marcel and Mathiesen, Joachim and Renard, Fran{\c{c}}ois and Le Borgne, Tanguy},
  title   = {Dynamic Multiphase Flow Triggers Chaotic Mixing in Porous Media},
  journal = {Proceedings of the National Academy of Sciences of the United States of America},
  volume  = {123},
  number  = {34},
  pages   = {e2612884123},
  year    = {2026},
  doi     = {10.1073/pnas.2612884123}
}

@article{Hyman2014,
  author  = {Hyman, Jeffrey D. and Winter, C. Larrabee},
  title   = {Stochastic Generation of Explicit Pore Structures by Thresholding Gaussian Random Fields},
  journal = {Journal of Computational Physics},
  volume  = {277},
  pages   = {16--31},
  year    = {2014},
  doi     = {10.1016/j.jcp.2014.07.046}
}

@article{Buchgraber2012,
  author  = {Buchgraber, Markus and Al-Dossary, M. and Ross, C. M. and Kovscek, A. R.},
  title   = {Creation of a Dual-Porosity Micromodel for Pore-Level Visualization of Multiphase Flow},
  journal = {Journal of Petroleum Science and Engineering},
  volume  = {86--87},
  pages   = {27--38},
  year    = {2012},
  doi     = {10.1016/j.petrol.2012.03.012}
}

@article{Bordoloi2022,
  author  = {Bordoloi, Ankur Deep and Scheidweiler, David and Dentz, Marco and Bouabdellaoui, Mohammed and Abbarchi, Marco and de Anna, Pietro},
  title   = {Structure Induced Laminar Vortices Control Anomalous Dispersion in Porous Media},
  journal = {Nature Communications},
  volume  = {13},
  pages   = {3820},
  year    = {2022},
  doi     = {10.1038/s41467-022-31552-5}
}

@article{Haefeli2014,
  author  = {H{\"a}feli, Richard and Altheimer, Marco and Butscher, Denis and Rudolf von Rohr, Philipp},
  title   = {PIV Study of Flow through Porous Structure Using Refractive Index Matching},
  journal = {Experiments in Fluids},
  volume  = {55},
  pages   = {1717},
  year    = {2014},
  doi     = {10.1007/s00348-014-1717-5}
}

@article{Winkler2020,
  author  = {Winkler, Mathias and Gjennestad, Magnus Aa. and
             Bedeaux, Dick and Kjelstrup, Signe and
             Cabriolu, Raffaela and Hansen, Alex},
  title   = {Onsager-Symmetry Obeyed in Athermal Mesoscopic Systems:
             Two-Phase Flow in Porous Media},
  journal = {Frontiers in Physics},
  volume  = {8},
  pages   = {60},
  year    = {2020},
  doi     = {10.3389/fphy.2020.00060}
}

@article{Berg2026,
  author  = {Berg, Steffen and Armstrong, Ryan T. and
             R{\"u}cker, Maja and Hansen, Alex and
             Kjelstrup, Signe and Bedeaux, Dick},
  title   = {From interface dynamics to Darcy scale description
             of multiphase flow in porous media},
  journal = {Advances in Colloid and Interface Science},
  volume  = {351},
  pages   = {103791},
  year    = {2026},
  doi     = {10.1016/j.cis.2026.103791}
}

@article{Christopher2024,
  author  = {Christopher, Jay and Rooney, Liam M. and Donnachie, Mark
             and Uttamchandani, Deepak and McConnell, Gail and Bauer, Ralf},
  title   = {Low-cost 3D printed lenses for brightfield and fluorescence microscopy},
  journal = {Biomedical Optics Express},
  volume  = {15},
  number  = {4},
  pages   = {2224--2237},
  year    = {2024},
  doi     = {10.1364/BOE.514653}
}

@article{Armstrong2016,
  author  = {Armstrong, Ryan T. and McClure, James E. and
             Berrill, Mark A. and R{\"u}cker, Maja and
             Schl{\"u}ter, Steffen and Berg, Steffen},
  title   = {Beyond Darcy's law: The role of phase topology and
             ganglion dynamics for two-fluid flow},
  journal = {Physical Review E},
  volume  = {94},
  number  = {4},
  pages   = {043113},
  year    = {2016},
  doi     = {10.1103/PhysRevE.94.043113}
}

@book{bluntMultiphaseFlowPermeable2017,
    address = {Cambridge},
    title = {Multiphase {Flow} in {Permeable} {Media}: {A} {Pore}-{Scale} {Perspective}},
    isbn = {978-1-107-09346-1},
    shorttitle = {Multiphase {Flow} in {Permeable} {Media}},
    url = {https://www.cambridge.org/core/books/multiphase-flow-in-permeable-media/455B5CA44AAF69F06A82DE091CD631C7},
    doi = {10.1017/9781316145098},
    urldate = {2025-11-21},
    publisher = {Cambridge University Press},
    author = {Blunt, Martin J.},
    year = {2017},
}

@article{macdonaldComparingMicrofluidicPerformance2017,
    title = {Comparing {Microfluidic} {Performance} of {Three}-{Dimensional} ({3D}) {Printing} {Platforms}},
    volume = {89},
    issn = {0003-2700},
    url = {https://doi.org/10.1021/acs.analchem.7b00136},
    doi = {10.1021/acs.analchem.7b00136},
    number = {7},
    urldate = {2026-09-22},
    journal = {Analytical Chemistry},
    author = {Macdonald, Niall P. and Cabot, Joan M. and Smejkal, Petr and Guijt, Rosanne M. and Paull, Brett and Breadmore, Michael C.},
    month = apr,
    year = {2017},
    pages = {3858--3866},
}

@article{Reynoso2021,
  author  = {Reynoso, Mateo and Gauli, Ishan and Measor, Philip},
  title   = {Refractive index and dispersion of transparent
             3D printing photoresins},
  journal = {Optical Materials Express},
  volume  = {11},
  number  = {10},
  pages   = {3392--3397},
  year    = {2021},
  doi     = {10.1364/OME.438040}
}

@inproceedings{Rank2019,
  author    = {Rank, Manuel and Heinrich, Andreas},
  title     = {Measurement and use of the refractive index distribution
               in optical elements created by additive manufacturing},
  booktitle = {Advanced Fabrication Technologies for Micro/Nano Optics
               and Photonics XII},
  volume    = {10930},
  year      = {2019},
  publisher = {SPIE},
  doi       = {10.1117/12.2507262}
}

@article{Miner2026,
  author  = {Miner, Dallin S. and Viglione, Matthew S. and Hooper, Kent and Woolley, Adam T. and Nordin, Gregory P.},
  title   = {Fast multi-resolution 3D printing of microfluidics: enabling 2~$\mu$m channels and ultra-compact mixers},
  journal = {Microsystems \& Nanoengineering},
  volume  = {12},
  pages   = {66},
  year    = {2026},
  doi     = {10.1038/s41378-026-01194-4}
}

@article{Kawale2019,
  author  = {Kawale, Durgesh and Jayaraman, Jishnu and Boukany, Pouyan E.},
  title   = {Microfluidic rectifier for polymer solutions flowing through porous media},
  journal = {Biomicrofluidics},
  volume  = {13},
  number  = {1},
  pages   = {014111},
  year    = {2019},
  doi     = {10.1063/1.5050201}
}

@article{Vezies2026,
  author  = {Vezies, Cl{\'e}ment and Zavattieri, Pablo and Silin, Nicol{\'a}s},
  title   = {A Non-Reciprocal Architected Porous Medium},
  journal = {Advanced Functional Materials},
  volume  = {36},
  number  = {47},
  pages   = {e75486},
  year    = {2026},
  doi     = {10.1002/adfm.75486}
}

@article{Crandall2008,
  author  = {Crandall, Dustin and Ahmadi, Goodarz and Leonard, Douglas and Ferer, Martin and Smith, Duane H.},
  title   = {A new stereolithography experimental porous flow device},
  journal = {Review of Scientific Instruments},
  volume  = {79},
  number  = {4},
  pages   = {044501},
  year    = {2008},
  doi     = {10.1063/1.2903740}
}

\section*{Statements and Declarations}

\subsection*{Author contributions}

Y.D.: Conceptualisation, methodology, investigation, software, formal analysis, data curation, visualisation, and writing -- original draft. 
J.K.P.: Formal analysis, interpretation of results, and writing -- review and editing. 
J.D.: Conceptualisation, methodology, co-development of the experimental setup, development of visualisation methods, and writing -- review and editing. 
E.E.: Conceptualisation, supervision, project administration, and writing -- review and editing. 

\subsection*{Funding}

This project has received funding from the European Union's Horizon Europe research and innovation programme under the Marie Skłodowska-Curie Doctoral Network FLUXIONIC, grant agreement No. 101119598. We also acknowledge the support of the Research Council of Norway through the Centres of Excellence funding scheme, project No.
262644 (PoreLab), the FRINATEK program, Project No. 325819 (Mixing in Multiphase flow through Microporous Media), and the INTPART program, Project No. RCN 309139 (Non-Newtonian Flow in Porous Media).

\subsection*{Competing interests}

The authors declare that they have no competing interests.

\subsection*{Data availability}

The complete materials and equipment list, processed data, analysis scripts, design files, LabKit classifiers, and representative image data supporting the findings of this study are available in the accompanying Zenodo
repository~\citep{Dumay2026Zenodo}.
The full raw image data are not included in the repository because of their large file size, but will be made available by the corresponding author upon reasonable request.

\end{document}